\documentclass[pra,aps,groupedaddress,floatfix,twocolumn,superscriptaddress]{revtex4}
\usepackage{amsmath,amssymb,multirow,epsfig,bm,pifont}
\usepackage[linkcolor=black,citecolor=black,urlcolor=blue,colorlinks=true,linktocpage=true]{hyperref}
\usepackage{graphicx}
\usepackage{epsfig}
\usepackage{bm}
\usepackage{tabularx}
\usepackage{makecell}

\usepackage{feynmp-auto}

\newcommand{\beq}{\begin{equation}}
\newcommand{\eeq}{\end{equation}}
\newcommand{\bea}{\begin{eqnarray}}
\newcommand{\eea}{\end{eqnarray}}
\newcommand{\Tr}{\mathrm{Tr}}
\newcommand{\tr}{\mathrm{tr}}

\newcommand{\bd}{\boldsymbol}

\newcommand{\bra}[1]{\langle #1 |}
\newcommand{\ket}[1]{| #1 \rangle}
\newcommand{\bk}[2]{\langle #1 | #2 \rangle}

\begin{document}

\title{The entanglement spectrum and R\'enyi entropies of non-relativistic conformal fermions}

\author{William J. Porter}
\email{wjporter@live.unc.edu}
\affiliation{Department of Physics and Astronomy, University of North Carolina,
Chapel Hill, North Carolina, 27599-3255, USA}

\author{Joaqu\'{\i}n E. Drut}
\email{drut@email.unc.edu}
\affiliation{Department of Physics and Astronomy, University of North Carolina,
Chapel Hill, North Carolina, 27599-3255, USA}

\begin{abstract}
We characterize non-perturbatively the
R\'enyi entropies of degree $n=2,3,4$, and $5$ of three-dimensional, strongly coupled 
many-fermion systems in the scale-invariant regime of short interaction range and large scattering length, 
i.e. in the unitary limit. We carry out our calculations using lattice methods devised recently by us.
Our results show the effect of strong pairing correlations on the entanglement
entropy, which modify the sub-leading behavior for large subsystem sizes (as characterized by the dimensionless parameter $x=k_F L_A$,
where $k_F$ is the Fermi momentum and $L_A$ the linear subsystem size), but leave the leading order unchanged relative
to the non-interacting case. Moreover, we find that the onset of the sub-leading asymptotic regime 
is at surprisingly small $x\simeq 2 - 4$.
We provide further insight into the entanglement properties of this system by
analyzing the spectrum of the entanglement Hamiltonian of the two-body problem from weak to strong coupling.
The low-lying entanglement spectrum displays clear features as the strength of the coupling is varied, such as 
eigenvalue crossing and merging, a sharp change in the Schmidt gap, and scale invariance at unitarity. 
Beyond the low-lying component, the spectrum appears as a quasi-continuum distribution,
for which we present a statistical characterization; we find, in particular, that the mean shifts to infinity as the coupling is
turned off, which indicates that that part of the spectrum represents non-perturbative contributions to the 
entanglement Hamiltonian.
In contrast, the low-lying entanglement spectrum evolves to finite values in the noninteracting limit.
The scale invariance of the unitary regime guarantees that our results are universal features intrinsic
to three-dimensional quantum mechanics and represent a well-defined prediction for ultracold atom 
experiments, which were recently shown to have direct access to the entanglement entropy.
\end{abstract}

\date{\today}
\maketitle

\section{Introduction}

This is an incredibly exciting time for research in ultracold atomic physics. The degree of control that experimentalists
have achieved continues to rise, year after year, along with their ability to measure collective properties in
progressively more ingenious ways (see e.g.~\cite{UltracoldBook, UltracoldRMP1, UltracoldRMP2, UltracoldLattices1}). 
Indeed, after the realization of Bose-Einstein condensates over two decades ago~\cite{BEC1,BEC2,BEC3} (see also~\cite{BECPhysicsFocus}), 
followed by Fermi condensates in 2004~\cite{FermionCondensate1}, the field entered an accelerated phase 
and rapidly developed control of multiple parameters such as temperature, polarization, and interaction strength 
(in alkali gases via Feshbach resonances, see e.g.~\cite{ExpReview}, and more recently in alkaline-earth gases via orbital 
resonances, see e.g.~\cite{OrbitalResonances1,OrbitalResonances2,OrbitalResonances3}),
as well as exquisite tuning of external trapping potentials. Additionally, multiple properties can be measured, ranging from
the equation of state (see e.g.~\cite{EoS1,EoS2,EoSViewPoint}) to hydrodynamic response (see e.g.~\cite{JET1,JET2}) and, more recently, the entanglement entropy~\cite{Greiner1,Greiner2}.

This sustained progress has strengthened the intersections with other areas of physics, in particular
modern condensed matter physics and quantum information~\cite{CondMatQI}, as well as with nuclear~\cite{UFGBook} and 
particle physics~\cite{QCDQEDsimulationWatoms1,QCDQEDsimulationWatoms2,QCDQEDsimulationWatoms3}. Quantum simulation by fine manipulation of nuclear 
spins, electronic states, and optical lattices, now appears more realistic than ever~\cite{QuantumSimulation1,QuantumSimulation2,QuantumSimulation3}.
At the interface between many of those areas lies a deceptively simple non-relativistic scale invariant system:
the unitary Fermi gas, which corresponds to the limit of vanishing interaction range $r_0$ and infinite s-wave scattering
length $a$, i.e.
\beq
0 \leftarrow r_0 \ll n^{-1} \ll a \to \infty
\eeq
where $n$ is the density; this regime corresponds to the threshold of two-body bound-state formation.

Both a model for dilute neutron matter and an actually realized resonant atomic gas,
this universal spin-$1/2$ system has brought together the nuclear~\cite{UFGNP0, UFGNP1, UFGNP2}, 
atomic~\cite{UFGAMO}, and condensed matter physics areas~\cite{UFGCondMat1,UFGCondMat2,UFGCondMat3}, 
as well as the AdS/CFT area~\cite{UFGAdSCFT1, UFGAdSCFT2, UFGAdSCFT3},
due to the underlying non-relativistic conformal invariance~\cite{SonNishida}. While many properties of this 
quintessential many-body problem are known (see e.g.~\cite{UFGBook} for an extensive review), other properties like entanglement and quantum 
information aspects have thus far remained unexplored, which brings us to our main point.

As this work is being written, quantum information concepts are increasingly becoming part of the modern language of quantum 
many-body physics (see e.g. Refs.~\cite{RevModPhys1,RevModPhys2,RevModPhys3,CondMatQI}), in particular with regards to 
the characterization of topological phases of matter and quantum computation, but also in connection with black holes (see e.g.~\cite{Srednicki}) and 
string theory (see e.g.~\cite{StringQI}).
In the past decade or so, a large body of work has been produced characterizing the entanglement properties of low dimensional systems 
(especially those with spin degrees of freedom~\cite{Melko1,Melko2,Melko3}) at quantum phase transitions (in particular those with topological
order parameters that defy a local description) as well as systems of noninteracting fermions and bosons~\cite{NonIntS1,NonIntS2,NonIntS3,NonIntS4,NonIntS5}, which
presented a challenge of their own.

With that new perspective in mind, in this work we set out to characterize the entanglement properties of the unitary Fermi gas
using non-perturbative lattice methods.
We analyze the reduced density matrix, entanglement spectrum, and associated R\'enyi entanglement entropies of the two-body problem 
by implementing an exact projection technique on the lattice.
For the many-body problem, we use a Monte Carlo method developed by us in Refs.~\cite{DrutPorter1,DrutPorter2},
based on the work of Ref.~\cite{Grover}, to calculate the $n$-th R\'enyi entanglement entropy. 
We showed in that work that our method overcomes the signal-to-noise problem of na\"\i ve Monte Carlo approaches.
We did that using the 1D Fermi-Hubbard model as a test case, but to our knowledge no previous calculations have 
been attempted for the challenging case of 3D Fermi gases.

The remainder of this paper is organized as follows: In Sec.~\ref{sec:Defs} we present the
main definitions and set the stage for Sec.~\ref{sec:Method}, where we explain how we 
carry out our calculations of the entanglement spectrum and entanglement entropies in
two- and many-fermion systems. For completeness, we also include in that section
a discussion on how to avoid the signal-to-noise issue that plagues entanglement-entropy 
calculations in the many-body case. We extend that discussion to the case of 
bosons in the same section. In Sec.~\ref{sec:Results1} we show 
our results for the entanglement spectrum and entropies of the two-body system along the
BCS-BEC crossover, and in Sec.~\ref{sec:Results2} we present the R\'enyi entanglement entropies of 
many fermions at unitarity. We present a summary and our main conclusions in Sec.~\ref{sec:Conclusions}.
The appendices contain more detailed explanations of our few- and many-body methods.

\begin{figure}[t]
\includegraphics[width=0.9\columnwidth]{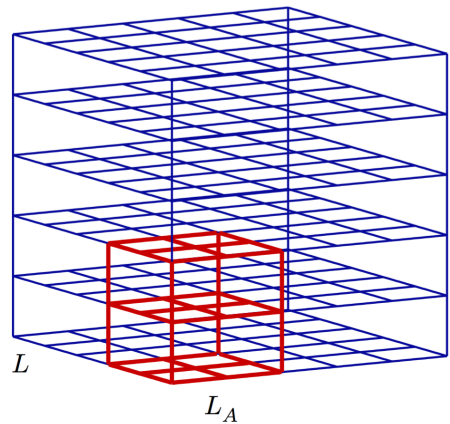}
\caption{\label{Fig:Region}
The (bipartite) entanglement entropies computed in this work correspond to partitioning the system
into a subsystem $A$ (in coordinate space, but it can also be defined in momentum space) 
and its complement $\bar A$. In practice, the calculations are carried out on systems that live in a cubic 
lattice of side $L$, and the subsystems are defined by cubic subregions of side $L_A \leq L$. The reduced 
density matrix $\hat \rho^{}_A$ of the open system $A$ contains the information about entanglement between $A$ 
and $\bar A$, and is obtained by tracing the full density matrix over the states supported by $\bar A$, which form 
the Hilbert space $\mathcal H_{\bar A}$.
}
\end{figure}
%

\section{Definitions: Hamiltonian, density matrices, and the entanglement entropy~\label{sec:Defs}}

The Hamiltonian governing the dynamics of resonant fermions can be written as
\beq
\hat{H} = \hat{T} + \hat{V},
\eeq
where the non-relativistic kinetic energy operator is
\beq
\hat{T} = \sum_{s = \uparrow,\downarrow} \int d^3 r\;\hat{\psi}^{\dagger}_{s}({\bf r})\left(-\frac{\nabla^2}{2m}\right)\hat{\psi}^{}_{s}({\bf r}),
\eeq
where $\hat{\psi}^{\dagger}_{s}({\bf r})$ and $\hat{\psi}^{}_{s}({\bf r})$ are the creation and annihilation operators of particles of
spin $s=\uparrow,\downarrow$ at location ${\bf r}$.

The two-body, zero-range interaction operator is
\beq
\hat{V} = -g \int d^3 r \; \hat{\psi}^{\dagger}_{\uparrow}({\bf r})\hat{\psi}^{}_{\uparrow}({\bf r})
\hat{\psi}^{\dagger}_{\downarrow}({\bf r})\hat{\psi}^{}_{\downarrow}({\bf r}),
\eeq
where the bare coupling $g$ is tuned to the desired physical situation. By definition, the limit of unitarity is achieved by
requiring that the ground state of the two-body problem lies at the threshold of bound-state formation 
(note that in 1D and 2D bound states form at arbitrarily small attractive coupling, but a finite value is required in 3D). 
Because our work was carried out in a finite volume with periodic boundary
conditions, we used L\"uscher's formalism~\cite{Luescher1,Luescher2} to relate the bare coupling to the scattering length in the analysis
of the BCS-BEC crossover. We describe that procedure below, when showing the results for the two-body problem.

The full, normalized density matrix of the system is 
\beq
\hat{\rho} = \frac{e^{-\beta\hat H}}{\mathcal Q},
\eeq
where
\beq
\mathcal{Q} = \Tr^{}_{\mathcal H} \left[ e^{-\beta\hat H} \right],
\eeq
is of course the canonical partition function, and $\mathcal H$ is the full Hilbert space.
In this work we are concerned with systems in a pure state, namely the ground state $\ket{\Xi}$, 
such that the full density matrix can be written as
\beq
\hat{\rho} = \ket{\Xi}\bra{\Xi}.
\eeq
Both in the few- and many-body systems we analyze here, the ground-state density matrix will
be approached by a projection method we describe below.

\begin{figure}[t]
\includegraphics[width=0.97\columnwidth]{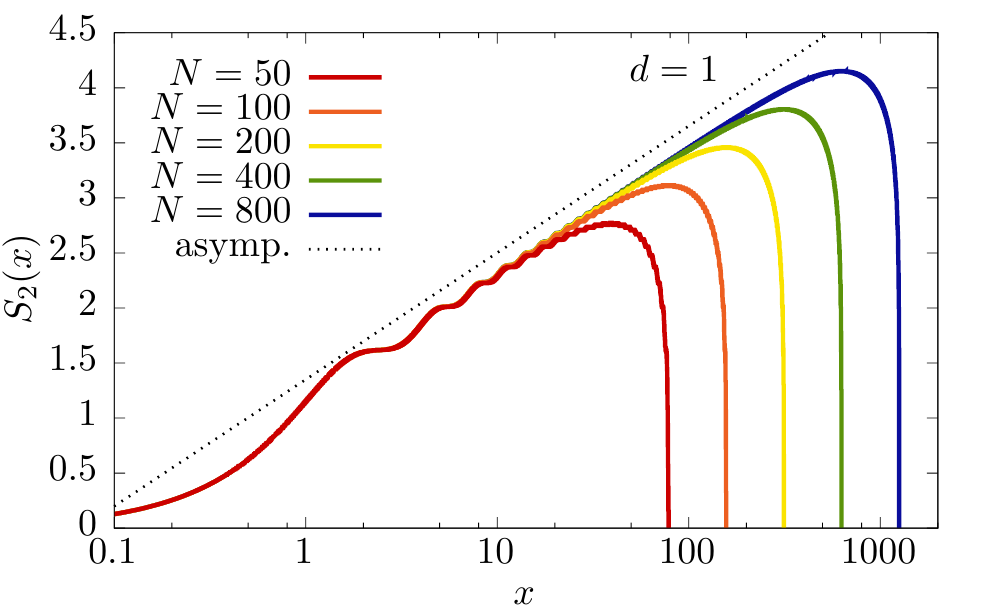}
\includegraphics[width=0.97\columnwidth]{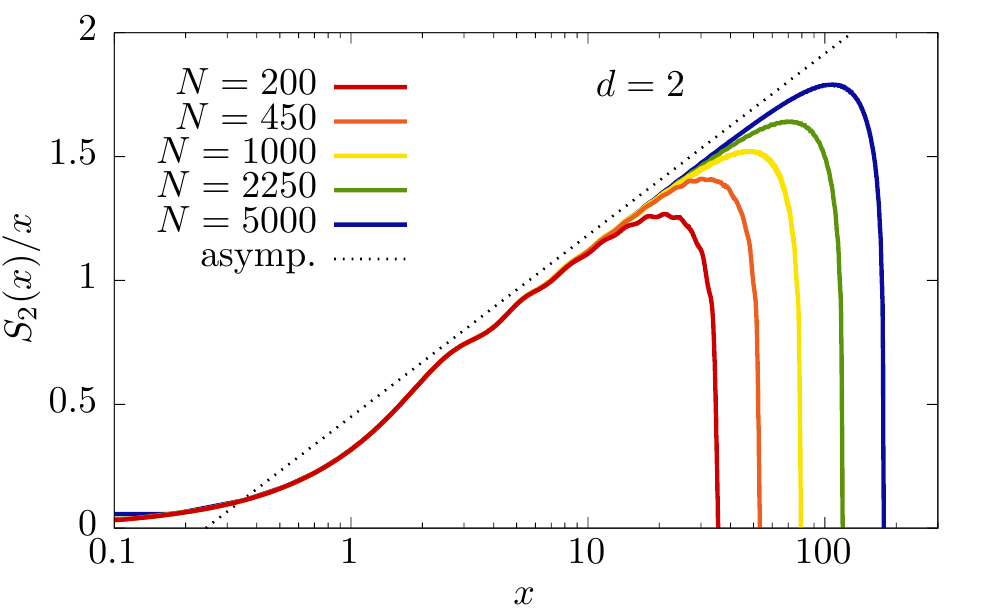}
\includegraphics[width=0.97\columnwidth]{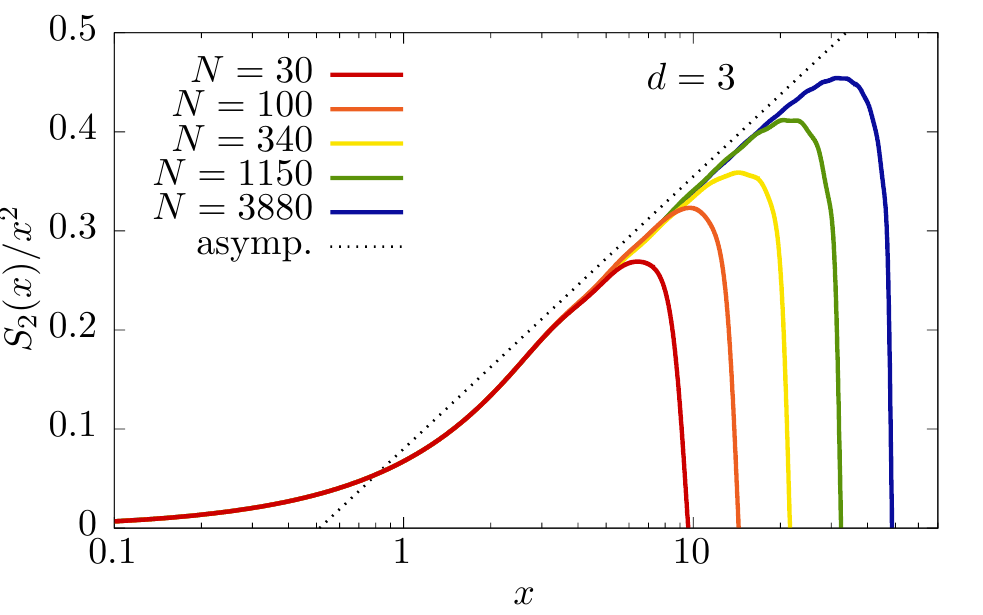}
\caption{\label{Fig:scalingFG} Second R\'enyi entropy $S_2$ of $N$ non-interacting fermions
in $d=1,2,3$ dimensions (top to bottom) as a function of $x=k_F L_A$, where $A$ 
is a segment, square, and cubic region, respectively, and $L_A$ is the corresponding linear size; $k_F$ is the
Fermi momentum. $S_2$ is scaled by the surface area dependence, namely $x$ 
and $x^2$ in 2D and 3D, respectively. The $x$ axis is plotted logarithmically to show that, up to finite-size effects,
the results heal to the expected asymptotic regime of linear dependence with $\log_{10} x$ (dashed line).
This regime sets in at $x \simeq 2 - 4$ across all $d$.
Finite-size effects appear as a sudden drop at large $x$.
}
\end{figure}

A subsystem $A$ and its complement $\bar A$ (in coordinate or momentum space, see Fig.~\ref{Fig:Region}) 
support states that belong to Hilbert spaces
$\mathcal{H}^{}_{A}$ and $\mathcal{H}^{}_{\bar{A}}$, respectively, such that the Hilbert space $\mathcal H$
of the full system can be written as a direct product space
\beq
\mathcal{H} = \mathcal{H}^{}_{A} \otimes \mathcal{H}^{}_{\bar{A}}.
\eeq

The density matrix $\hat{\rho}^{}_{A}$ of subsystem $A$, usually referred to as the reduced density matrix, 
is defined by tracing over the degrees of freedom supported by $\bar A$, i.e. tracing over the states in 
$\mathcal{H}^{}_{\bar{A}}$:
\beq
\hat{\rho}^{}_{A} = \Tr_{\mathcal{H}^{}_{\bar{A}}}\hat{\rho}.
\eeq

Based on this definition, the properties of $A$ as an open subsystem can be formulated and computed
using operators with support in $A$. In particular, a quantitative measure of entanglement between $A$ and 
$\bar A$ is given by the von Neumann entanglement entropy,
\beq
S^{}_{\mbox{vN},A} = -\Tr_{\mathcal{H}^{}_{A}}\left [ \hat{\rho}^{}_A\ln\hat{\rho}^{}_A\right ],
\eeq
and by the $n$-th order R\'enyi entanglement entropy,
\beq
S^{}_{n,A} = \frac{1}{1-n}\ln \Tr_{\mathcal{H}^{}_{A}}\left[ \hat{\rho}^{n}_{A} \right].
\eeq
Naturally, these entropies vanish when $A$ is the whole system, as then there is full knowledge of 
the state of the system. In any other case, the entanglement entropy will be non-zero, unless the
ground state factorizes into a state living in $A$ and a state living in $\bar A$.
Because the entanglement between $A$ and $\bar A$ happens across the boundary that separates
those regions, it is natural to expect $S^{}_{n,A}$ to be extensive with the size of that boundary,
i.e. proportional to the area delimiting $A$. This point was the topic of many papers in the last decade or so,
especially in connection with quantum phase transitions (see e.g.~\cite{EEandQPT}).

It was rigorously shown in recent years, however, that the R\'enyi entropy of non-interacting fermions
with a well-defined Fermi surface presents a logarithmic violation of the area law~\cite{NonIntS1,NonIntS2,NonIntS3,NonIntS4,NonIntS5}. 
This abnormality was confirmed numerically with the aid of overlap-matrix methods~\cite{OverlapMatrixMethod}, which we reproduce in Fig.~\ref{Fig:scalingFG}, where
we explicitly show said logarithmic dependence (dashed line) as a function of $x=k_F L_A$, where $k_F$ is the
Fermi momentum and $L_A$ is the linear size of region $A$, such that 
\bea
x = k_F L_A &=& \frac{\pi N}{2} \frac{L_A}{L}  \ \ \ \ \ \  \ \ \text{in 1D,} \\
&=& ({2 \pi N})^{1/2} \frac{L_A}{L} \ \ \ \ \ \text{in 2D,} \\
&=& (3 \pi^2 N)^{1/3} \frac{L_A}{L}\ \ \ \ \ \text{in 3D},
\eea
where $N$ is the total particle number. Note that, at large enough $x$, finite size 
effects eventually take over and the entropy quickly tends to zero. The sub-leading oscillations were studied in detail
in Ref.~\cite{Oscillations1}.

Although resonant fermions are strongly coupled (the regime is non-perturbative and away from
any regime with small dimensionless parameters), we can expect $S^{}_{n,A}$ to follow a similar trend as the 
non-interacting gas, for the following reasons.
First, resonant fermions have a distinguishable Fermi surface (note, however, that that is quickly 
lost as one proceeds towards the BEC side of the resonance), whose role in the entanglement entropy has been
emphasized many times (see e.g.~\cite{Swingle}). Second, our experience with $S^{}_{n,A}$ for the 
Hubbard model in other cases~\cite{DrutPorter1} indicates that very strong couplings $U/t$ are needed
even in 1D (where quantum fluctuations are qualitatively stronger than in 3D) in order for $S^{}_{n,A}$ to noticeably 
depart from the non-interacting result. Thus, we anticipate a similar behavior for resonant fermions as that of the bottom 
panel of Fig.~\ref{Fig:scalingFG}; the latter provides some qualitative knowledge of where the leading logarithmic and 
sub-leading dependence sets in for $S^{}_{n,A}$ as a function of $x = k_F L_A$. In fact, as we will see below,
the onset of the asymptotic behavior (meaning dominated by leading and sub-leading dependence on $x$) at $x \simeq 2 - 4$ is the 
same for unitarity as for the non-interacting case. This is surprising, as there is no obvious reason for that to be the case:
had this onset appeared at $x\simeq 10$, the calculations in this work would not have been possible, as they would
have required huge lattices. We return to this discussion below, when presenting our results for the many-body case.

\section{Method~\label{sec:Method}}

In this section we explain the two approaches used in this work.
We address the two-body problem first, which we solved with a direct (i.e. non-stochastic) 
projection method on the lattice. This problem can be solved exactly by changing to center-of-mass and relative coordinates. 
However, doing so implies using a method that only works in that case, and we are interested in techniques that
can be used in a variety of situations (e.g. in the presence of external fields, more than
two particles, time-dependent cases, and so forth). We then address the many-body problem 
using a method recently put forward by us, which we first presented and tested for 
one-dimensional systems in Ref.~\cite{DrutPorter2}.

Although both approaches make use of an auxiliary field transformation, the ultimate utility of this technique is markedly different in 
each case.  We detail below the portion of the formalism common to both approaches, treating in subsequent sections the details 
of their divergence from common assumptions and notation.

At chemical potential $\mu$ and inverse temperature $\beta$, the grand canonical partition function $\mathcal Z$ is defined via
\beq
\mathcal Z = \Tr \,\left[ e^{-\beta (\hat{H} - \mu \hat{N})} \right]
\eeq
for Hamiltonian $\hat{H}$ and particle-number operator $\hat{N}$.  Writing the inverse temperature as an integer number $N^{}_{\tau}$ of steps, we implement a symmetric Suzuki-Trotter decomposition with the goal of separating each operator into distinct one- and two-body factors.  For the Boltzmann factor, we obtain
\bea
\label{Eq:TS}
e^{-\beta (\hat{H} - \mu \hat{N})} &=& \prod_{j=1}^{N^{}_{\tau}}e^{-\tau\hat{K}/2}e^{-\tau\hat{V}}e^{-\tau\hat{K}/2}+\mathcal{O}(\tau^2)
\eea
were we define 
\beq
\hat{K} = \hat{T}-\mu\hat{N}.
\eeq  

At each position ${\bf r}$ and for each of the $N^{}_{\tau}$ factors, we decompose the interaction via the introduction of a Hubbard-Stratonovich auxiliary field $\sigma$ which we choose to be of a continuous and compact form~\cite{MCReviews2,MCReviews4}.  More specifically for each spacetime position $({\bf r},\tau^{}_{j})$, where ${\bf r} \in [0,L)^3$ and $\tau^{}_{j} = j \tau$ for some $1\le j \le N^{}_{\tau}$, we write
\begin{figure}[t]
\includegraphics[width=1.0\columnwidth]{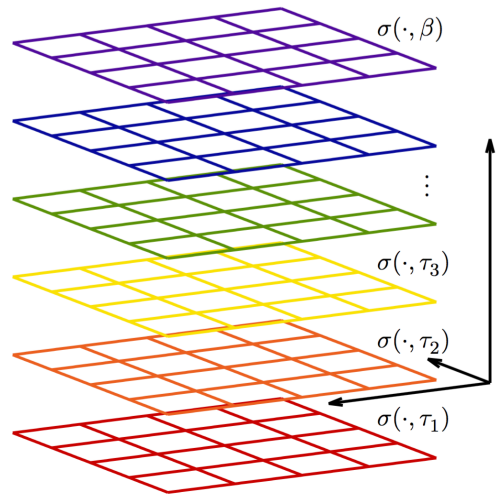}
\caption{\label{Fig:spacetime}
Shown here is a representation of the lattice used in our calculations.
Each horizontal lattice plane represents the 3D lattice where the system lives, and the vertical stacking of the planes
represents the imaginary time direction. Although the original Hamiltonian is time-independent, the auxiliary field
$\sigma$ that represents the interaction is supported by a spacetime lattice and induces a time dependence that 
disappears upon averaging.}
\end{figure}
\bea
\label{Eq:HS}
e^{\tau g\hat{n}^{}_{\uparrow}\hat{n}^{}_{\downarrow}}\!=\!
\int_{-\pi}^{\pi} \frac{d\sigma}{2 \pi}  
\left( \openone +B \; \hat{n}^{}_{\uparrow}\sin\sigma \right) \! 
\left( \openone +B \; \hat{n}^{}_{\downarrow}\sin\sigma\right)
\eea
having suppressed the spacetime dependence of the field $\sigma$ and the spatial dependence of the fermion density operators 
$\hat{n}^{}_{s}({\bf r}) = \hat{\psi}^{\dagger}_{s}({\bf r})\hat{\psi}^{}_{s}({\bf r})$ where $s = \uparrow,\downarrow$.  Knowing that 
$\hat{n}^{}_{s}({\bf r})$ is idempotent, it follows that
\beq
e^{\tau g\hat{n}^{}_{\uparrow}\hat{n}^{}_{\downarrow}} = 1 + (e^{\tau g}-1)\hat{n}^{}_{\uparrow}\hat{n}^{}_{\downarrow},
\eeq
which shows that the constant $B$ satisfies 
\beq
e^{\tau g} -1 = \frac{B^2}{2}.  
\eeq
Collecting the integration measures, we obtain a path-integral form of the partition function accurate to quadratic order in the temporal lattice spacing, writing
\beq
\mathcal Z = \int \mathcal D\sigma \; \Tr\;\hat{\mathcal U}[\sigma]+\mathcal{O}(\tau^2)
\eeq
where
\beq
\label{Eq:OpU}
\hat{\mathcal U}[\sigma] = \prod_{j=1}^{N^{}_{\tau}}\hat{\mathcal U}^{}_{j}[\sigma],
\eeq
and the individual factors are
\bea
\hat{\mathcal U}^{}_{j}[\sigma] = e^{-\tau\hat{K}/2}\prod_{{\bf r}}\left( \openone +B \; \hat{n}^{}_{\uparrow}({\bf r})\sin\sigma({\bf r},\tau^{}_{j}) \right)\\
\times\left( \openone +B \; \hat{n}^{}_{\downarrow}({\bf r})\sin\sigma({\bf r},\tau^{}_{j})\right)e^{-\tau\hat{K}/2}.
\eea

As the kinetic energy operator $\hat{T}$ and the number operator $\hat{N}$ are already written as products of flavor-specific operators, we may partition the operator $\hat{\mathcal U}$ into individual factors each of which assumes responsibility for the evolution of a particular fermion species $s = \uparrow,\downarrow$.  
We do this by defining operators $\hat{T}^{}_{s}$, $\hat{N}^{}_{s}$, and $\hat{K}^{}_{s}$ for $s = \uparrow,\downarrow$ by
\beq
\hat{T}^{}_{s} = \int d^3 r\;\hat{\psi}^{\dagger}_{s}({\bf r})\left(-\frac{\nabla^2}{2m}\right)\hat{\psi}^{}_{s}({\bf r}),
\eeq
\beq
\hat{N}^{}_{s} = \int d^3 r\;\hat{\psi}^{\dagger}_{s}({\bf r})\hat{\psi}^{}_{s}({\bf r}),
\eeq
and $\hat{K}^{}_{s} = \hat{T}^{}_{s} - \mu \hat{N}^{}_{s}$. We then write
\bea
\hat{\mathcal U}^{}_{j,s}[\sigma] = e^{-\tau\hat{K}^{}_{s}/2}\prod_{{\bf r}}&\!\!\!\left( \openone +B \; \hat{n}^{}_{s}({\bf r})\sin\sigma({\bf r},\tau^{}_{j})\right) \\
&\times \;e^{-\tau\hat{K}^{}_{s}/2}, \nonumber
\eea
such that
\beq
\hat{\mathcal U}[\sigma] = \hat{\mathcal U}^{}_{\uparrow}[\sigma]\;\hat{\mathcal U}^{}_{\downarrow}[\sigma],
\eeq
where
\beq
\hat{\mathcal U}^{}_{s}[\sigma] = \prod_{j=1}^{N^{}_{\tau}}\hat{\mathcal U}^{}_{j,s}[\sigma].
\eeq
Performing the required Fock-space trace, the exponential form of each factor in the above provides (see e.g.~\cite{MCReviews3})
\beq
\label{Eq:GeneralZ}
\mathcal Z = \int \mathcal D\sigma \; \det\left(\openone + \bd {U}^{}_{\uparrow}[\sigma]\right)\det\left(\openone + \bd{U}^{}_{\downarrow}[\sigma]\right),
\eeq
where we have suppressed higher-order contributions in $\tau$ (which are of order $\tau^2$), and written a matrix 
$\bd{U}^{}_{s}[\sigma]$ for the restriction of each of the operators $\hat{\mathcal U}^{}_{s}[\sigma]$ to the single-particle Hilbert space.  
Each of those matrices contains an overall factor of the fugacity 
\beq
z \equiv e^{\beta \mu}.
\eeq
In what follows, we exhibit this factor explicitly and redefine the matrices $\bd{U}^{}_{s}[\sigma]$ to 
reflect this revision. In this work, we exclusively treat unpolarized systems, and so we may treat the determinants as equivalent in 
derivations that follow by writing
\beq
\mathcal Z = \int \mathcal D\sigma \; {\det}^2\!\left(\openone + z\,\bd{U}^{}_{}[\sigma]\right),
\eeq
and neglecting to denote the spin degree of freedom wherever context precludes confusion.

\subsection{Direct lattice approach to the entanglement spectrum of the two-body problem}
\subsubsection{Identifying the transfer matrix}

In order to illustrate the details as well as the generality of our technique, we show the main steps here in broad strokes and
leave the details for Appendix~\ref{Appendix1}.

Using the above path-integral form of $\mathcal Z$, we first isolate the two-body sector.  From the finite-temperature partition function Eq.~(\ref{Eq:GeneralZ}), 
we may derive the conventional virial expansion in powers of the fugacity for each spin, which is given by
\beq
\mathcal Z = \sum_{N_\uparrow,N_\downarrow=0}^{\infty} z^{N_\uparrow}_{\uparrow}z^{N_\downarrow}_{\downarrow} \mathcal{Q}^{}_{N_\uparrow,N_\downarrow},
\eeq
where we have identified the coefficient of the $N_s$-th power of the fugacity as the $N_s$-particle canonical partition function $\mathcal{Q}^{}_{N_\uparrow,N_\downarrow}$.  
Expanding the path-integral expression for the grand canonical partition function, we find that in terms of the matrix $\bd {U}^{}_{}[\sigma]$, the $(1+1)$-particle partition function is
\beq
\mathcal{Q}^{}_{1,1} = \int \mathcal D\sigma \; {\tr}^2\,{\bd U}^{}_{}[\sigma].
\eeq
The path integral in $\mathcal{Q}^{}_{1,1}$ above can be evaluated directly in a way that 
elucidates the form of the two-body transfer matrix. To that end, we define a four-index object from 
which the above squared trace may be obtained by suitable index contraction:
\beq
R_{ab,cd} = \int \mathcal D \sigma \;{\bd U}^{}_{}[\sigma]_{ac}\;{\bd U}^{}_{}[\sigma]_{bd}.
\eeq
The same four-index object, with indices properly contracted to account for antisymmetry, can be used to analyze the $(2+0)$-particle case.

We next write out each of the matrices ${\bd U}^{}_{}[\sigma]$ in its product form; that is, we reintroduce Eq.~(\ref{Eq:OpU})
in matrix form:
\beq
\label{Eq:MatU}
{\bd U}[\sigma] = \prod_{j=1}^{N^{}_{\tau}}{\bd U}^{}_{j}[\sigma].
\eeq
For each contribution to the $N$-body transfer matrix, exactly $N$ factors of the matrix $\bd{U}^{}_{}[\sigma]$ appear, 
and as a result each temporal lattice point appears in the integrand $N$ times.  
Turning to the individual factors, we write each of the matrices $\bd{U}^{}_{j}[\sigma]$ in such a way as to exhibit the interaction.  That is, we write
\beq
{\bd U}^{}_{j}[\sigma] = {\bd T}{\bd V}^{}_{j}[\sigma]{\bd T},
\eeq
where
\beq
\left[ {\bd T} \right]^{}_{{\bf k}{\bf k}'}  = e^{-\tau k^2/2}\delta_{{\bf k},{\bf k}'},
\eeq
is the single-particle form of the kinetic energy operator defined above (in momentum space),
and the (position-space representation of the) auxiliary external potential operator has matrix elements
\beq
\left[{\bd V}^{}_{j}[\sigma]\right]_{{\bf r}{\bf r}'} = \left(1 + B\sin\sigma({\bf r},\tau^{}_{j})\right)\delta^{}_{{\bf r}{\bf r}'}.
\eeq
At this point, all matrix elements have been written out and can be shifted around as needed to carry out the path 
integral. The only non-zero results are obtained, of course, when an even number (in this $N=2$ case no more than 2) of 
fields $\sigma({\bf r},\tau^{}_{j})$ appear in the integrand for the same values of $({\bf r},\tau^{}_{j})$.

This undoing of the Hubbard-Stratonovich transformation may seem a cumbersome or convoluted way to proceed, but it is useful 
in that it mechanically generates the correct expression for the $N$-body partition function for {\it any} particle content simply by 
differentiation of the fermion determinants. Moreover, this is accomplished without the need to deal with operator algebra and is 
easily generalized to bosons. In the $2$-body case, in particular, the above procedure results in 
\beq
R^{}_{ac,bd} = \left [M^{N^{}_{\tau}}_{2} \right]^{}_{ac,bd},
\eeq
where we have naturally identified the transfer matrix in the two-particle subspace
\bea
[M^{}_{2}]^{}_{ac,bd} &=& {\mathcal K}_{ab} {\mathcal K}_{cd} + (e^{\tau g}-1) {\mathcal I}_{abcd},
\eea
and where
\bea
{\mathcal K}_{ij} &=& \sum_{p}{\bd T}_{i p}{\bd T}_{pj}, \\
{\mathcal I}_{ijkl} &=& \sum_{p}{\bd T}_{i p}{\bd T}_{pj }{\bd T}_{k p}{\bd T}_{pl}.
\eea

The form of the transfer matrix lends itself to a useful diagrammatic representation,
which we show for the two- and three-particle cases (the latter derived in Appendix~\ref{Appendix1})
in Eqs.~(\ref{Eq:M2}),~(\ref{Eq:M3a}) and~(\ref{Eq:M3b}).

\begin{widetext}

\hspace{0.5in}

    \unitlength = 1mm
    \begin{fmffile}{diag2}
    \begin{equation}\label{Eq:M2}
[M^{}_{2}]^{}_{ac,bd} \;\;\;\; = \;\;\;\;\;\;\;\;
      \parbox{20mm}{\begin{fmfgraph*}(20,15)
		\fmfleft{i0,i1,i2,i3}
		\fmfright{o0,o1,o2,o3}
		\fmf{phantom}{i0,o0}
		\fmf{fermion}{i1,o1}
		\fmf{fermion}{i2,o2}
		\fmf{phantom}{i3,o3}
		\fmflabel{$c$}{i1}
		\fmflabel{$d$}{o1}
		\fmflabel{$a$}{i2}
		\fmflabel{$b$}{o2}
	\end{fmfgraph*}}
	\;\;\;\;\;\; \;\;+ \;\;\;\; (e^{\tau g}-1) \;\;\;\;
    \parbox{20mm}{\begin{fmfgraph*}(20,15)
		\fmfleft{i1,i2}
		\fmfright{o1,o2}
		\fmf{fermion,tension=0.01}{i1,v1}
		\fmf{fermion,tension=0.01}{i2,v1}
		\fmf{fermion,tension=0.01}{v1,o1}
		\fmf{fermion,tension=0.01}{v1,o2}
		\fmfdot{v1}
		\fmflabel{$c$}{i1}
		\fmflabel{$d$}{o1}
		\fmflabel{$a$}{i2}
		\fmflabel{$b$}{o2}
	\end{fmfgraph*}}
    \end{equation}
    \end{fmffile}

\hspace{0.5in}

    \unitlength = 1mm
    \begin{fmffile}{diag3}
    \begin{equation}\label{Eq:M3a}
[M^{}_{3}]^{}_{abc,def} \;\;\;\; = \;\;\;\;\;\;\;\;
      \parbox{20mm}{\begin{fmfgraph*}(20,15)
		\fmfleft{i0,i1,i2,i3,i4}
		\fmfright{o0,o1,o2,o3,o4}
		\fmf{phantom}{i0,o0}
		\fmf{fermion}{i1,o1}
		\fmf{fermion}{i2,o2}
		\fmf{fermion}{i3,o3}
		\fmf{phantom}{i4,o4}
		\fmflabel{$c$}{i1}
		\fmflabel{$f$}{o1}
		\fmflabel{$b$}{i2}
		\fmflabel{$e$}{o2}
		\fmflabel{$a$}{i3}
		\fmflabel{$d$}{o3}
	\end{fmfgraph*}}
	\;\;\;\;\;\; \;\;+ \;\;\;\; (e^{\tau g}-1) \;\;\;\;\;\;
	      \parbox{20mm}{\begin{fmfgraph*}(20,15)
		\fmfleft{i1,i2,i3}
		\fmfright{o1,o2,o3}
		\fmf{fermion}{i1,v1}
		\fmf{fermion}{i2,v1}
		\fmf{fermion}{i3,v1}
		\fmf{fermion}{v1,o1}
		\fmf{fermion}{v1,o2}
		\fmf{fermion}{v1,o3}
		\fmfblob{.25w}{v1}
		\fmflabel{$c$}{i1}
		\fmflabel{$f$}{o1}
		\fmflabel{$b$}{i2}
		\fmflabel{$e$}{o2}
		\fmflabel{$a$}{i3}
		\fmflabel{$d$}{o3}
	\end{fmfgraph*}}
    \end{equation}
    \end{fmffile}
    
\hspace{0.5in}

    \begin{fmffile}{int3}
    \begin{equation}\label{Eq:M3b}
      \parbox{20mm}{\begin{fmfgraph*}(20,15)
		\fmfleft{i1,i2,i3}
		\fmfright{o1,o2,o3}
		\fmf{fermion}{i1,v1}
		\fmf{fermion}{i2,v1}
		\fmf{fermion}{i3,v1}
		\fmf{fermion}{v1,o1}
		\fmf{fermion}{v1,o2}
		\fmf{fermion}{v1,o3}
		\fmfblob{.25w}{v1}
		\fmflabel{$c$}{i1}
		\fmflabel{$f$}{o1}
		\fmflabel{$b$}{i2}
		\fmflabel{$e$}{o2}
		\fmflabel{$a$}{i3}
		\fmflabel{$d$}{o3}
	\end{fmfgraph*}}
 \;\;\;\; \;\;\;\;= \;\;\;\;\;\;\;\;
    \parbox{20mm}{\begin{fmfgraph*}(20,15)
		\fmfleft{i1,i2,i3}
		\fmfright{o1,o2,o3}
		\fmf{fermion}{i1,o1}
		\fmf{fermion,tension=1.0}{i2,v1}
		\fmf{fermion,tension=0.0}{i3,v1}
		\fmf{fermion,tension=1.0}{v1,o2}
		\fmf{fermion,tension=0.0}{v1,o3}
		\fmfdot{v1}
		\fmflabel{$c$}{i1}
		\fmflabel{$f$}{o1}
		\fmflabel{$b$}{i2}
		\fmflabel{$e$}{o2}
		\fmflabel{$a$}{i3}
		\fmflabel{$d$}{o3}
	\end{fmfgraph*}}
	\;\;\;\;\;\; \;\;+ \;\;\;\;\;\;\;\;
    \parbox{20mm}{\begin{fmfgraph*}(20,15)
		\fmfleft{i1,i2,i3}
		\fmfright{o1,o2,o3}
		\fmf{fermion,tension=1.0}{i1,v1}
		\fmf{fermion,left=.5,tension=1.0}{i2,o2}
		\fmf{fermion,tension=0.0}{i3,v1}
		\fmf{fermion,tension=1.0}{v1,o1}
		\fmf{fermion,tension=0.0}{v1,o3}
		\fmfdot{v1}
		\fmflabel{$c$}{i1}
		\fmflabel{$f$}{o1}
		\fmflabel{$b$}{i2}
		\fmflabel{$e$}{o2}
		\fmflabel{$a$}{i3}
		\fmflabel{$d$}{o3}
	\end{fmfgraph*}}
	\;\;\;\;\;\;\;\;+\;\;\;\;\;\;\;\;
    \parbox{20mm}{\begin{fmfgraph*}(20,15)
		\fmfleft{i1,i2,i3}
		\fmfright{o1,o2,o3}
		\fmf{fermion,tension=0.0}{i1,v1}
		\fmf{fermion,tension=1.0}{i2,v1}
		\fmf{fermion}{i3,o3}
		\fmf{fermion,tension=0.0}{v1,o1}
		\fmf{fermion,tension=1.0}{v1,o2}
		\fmfdot{v1}
		\fmflabel{$c$}{i1}
		\fmflabel{$f$}{o1}
		\fmflabel{$b$}{i2}
		\fmflabel{$e$}{o2}
		\fmflabel{$a$}{i3}
		\fmflabel{$d$}{o3}
	\end{fmfgraph*}}
    \end{equation}
    \end{fmffile}

\hspace{0.5in}

\end{widetext}

\subsubsection{Obtaining the ground state and the reduced density matrix}

Having identified the transfer matrix allows us to design a projection method to approach the ground state
by repeated application of $M^{}_{2}$. Proposing a guess state $\ket{\Xi^{}_{0}}$, we extract the true two-particle 
ground state $\ket{\Xi^{}_{}}$ via
\beq
M^{N^{}_{\tau}}_{2}\ket{\Xi^{}_{0}}\xrightarrow[]{N^{}_{\tau}\to\infty}\ket{\Xi^{}_{}}.
\eeq
In practice, we compute the position-space wavefunction $\xi(x^{}_{\uparrow},x^{}_{\downarrow}) = \bk{x^{}_{\uparrow},x^{}_{\downarrow}}{\Xi^{}_{}}$.
Wavefunction in hand, we compute the matrix elements of the full density matrix $\hat{\rho}^{}_{}$ as
\bea
\bra{x^{}_{\uparrow},x^{}_{\downarrow}}\,\hat{\rho}\,\ket{x'^{}_{\uparrow},x'^{}_{\downarrow}} &=& \bk{x^{}_{\uparrow},x^{}_{\downarrow}}{\Xi^{}_{}}\bk{\Xi^{}_{}}{x'^{}_{\uparrow},x'^{}_{\downarrow}} \\
&=& \xi^{*}(x'^{}_{\uparrow},x'^{}_{\downarrow})\,\xi(x^{}_{\uparrow},x^{}_{\downarrow}).
\eea

From these, the elements of the reduced density matrix $\hat{\rho}^{}_{A}$ can be obtained as well. Given two states 
$\ket{s},\ket{s'}\in \mathcal{H}^{}_{A}$ for the subregion $A$, each state being specified by choosing for each 
particle either a position in $A$ or in the complement $\bar A$, we compute
\beq
\bra{s}\,\hat{\rho}^{}_{A}\,\ket{s'} =  \sum_{\;\;\;\;\;a \in \mathcal A^{}_{ss'}} (\ket{s}\otimes\ket{a})^{\dagger}\,\hat{\rho}\,(\ket{s'}\otimes\ket{a}),
\eeq
where, at each fixed pair of two-particle states $s,s'$, the sum is taken over all states $\ket{a}\in\mathcal{H}^{}_{\bar A}$ such that the state $\ket{s}\otimes\ket{a}\in \mathcal{H}^{}_{A} \otimes \mathcal{H}^{}_{\bar{A}}$ (resp. $\ket{s'}\otimes\ket{a}\in \mathcal{H}^{}_{A} \otimes \mathcal{H}^{}_{\bar{A}}$) is consistent with the first (resp. second) index of the matrix element being evaluated. We have denoted this set as $\mathcal A^{}_{ss'}$. From this matrix, we compute the entanglement spectrum $\sigma(\hat{H}^{}_{A})$, that is the spectrum of the entanglement Hamiltonian defined 
\beq
\hat{\rho}^{}_{A} = e^{-\hat{H}^{}_{A}},
\eeq
as well as the von Neumann and R\'enyi entanglement entropies. 

\subsection{Lattice Monte Carlo approach to the many-body problem}

To address the many-body system, we implement the Monte Carlo version of the algorithm outlined above. The
output of this algorithm, however, is not the ground-state wavefunction but rather the expectation value of the desired observable
in a projected state. In our case, the observable is of course the entanglement entropy. To obtain it, crucial intermediate steps 
are required that go beyond conventional Monte Carlo approaches. We therefore outline the basic formalism first, and
then proceed to explain the additional steps required to calculate $S_{n,A}$.

\subsubsection{Basic formalism}

Beginning with a largely arbitrary many-body state $\ket{\Omega^{}_{0}}$, we evolve the state forward in imaginary time by an 
extent $\beta$ via
\beq
\ket{\Omega(\beta)} = e^{-\beta \hat H}\ket{\Omega^{}_0},
\eeq
For large imaginary times, we have
\beq
\ket{\Omega(\beta)}\xrightarrow[]{\beta\to\infty}\ket{\Omega},
\eeq
where $\ket{\Omega}$ is the true ground state provided that $\bk{\Omega^{}_{0}}{\Omega} \ne 0$.

For an operator $\hat{O}$, we may obtain the ground-state expectation value by studying the asymptotic behavior of the function 
\beq
\label{Eq:OGS}
O(\beta) = \frac{1}{Z(\beta)}\bra{\Omega(\beta/2)}\,\hat{O}\,\ket{\Omega(\beta/2)},
\eeq
with the zero-temperature normalization defined as
\beq
Z(\beta) = \bk{\Omega(\beta/2)}{\Omega(\beta/2)} = \bra{\Omega^{}_{0}}\,e^{-\beta\hat{H}}\,\ket{\Omega^{}_{0}}.
\eeq

As derived in detail earlier, we implement a symmetric factorization of the Boltzmann weight [c.f. Eq.~(\ref{Eq:TS})] 
in order to separate factors depending only on the one-body 
kinetic-energy operator from the significantly more complicated two-body potential-energy operator responsible for 
the effects of the interaction. Following this approximation, we again implement an auxiliary field transformation 
[c.f. Eq.~(\ref{Eq:HS})] to represent the interaction factor. This allows us to write the ground-
state estimator of Eq.~(\ref{Eq:OGS}) defined above in path integral form as
\beq
O(\beta) = \frac{1}{Z(\beta)}\int \mathcal D\sigma\; P^{}_{\beta}[\sigma]\;O^{}_{\beta}[\sigma],
\eeq
while simultaneously demonstrating that
\beq
Z(\beta) = \int \mathcal D\sigma\; P^{}_{\beta}[\sigma].
\eeq
We have identified a naturally emerging probability measure $P^{}_{\beta}[\sigma]$ computed as 
\beq
P^{}_{\beta}[\sigma] = \bra{\Omega^{}_{0}}\,\hat{\mathcal U}^{}_{\beta}[\sigma]\,\ket{\Omega^{}_{0}},
\eeq
with the operator $\hat{\mathcal U}^{}_{\beta}[\sigma]$ defined as in Eq.~(\ref{Eq:OpU}) (setting $\mu=0$ in the kinetic energy factor since particle number is 
fixed in this formalism). The integrand takes the form
\beq
O^{}_{\beta}[\sigma] = \frac{\bra{\Omega^{}_{0}}\,\hat{\mathcal U}^{}_{\beta/2}[\sigma] \,\hat{O}\,\hat{\mathcal U}^{}_{\beta/2}[\sigma] \,\ket{\Omega^{}_{0}}}{\bra{\Omega^{}_{0}}\,
\hat{\mathcal U}^{}_{\beta}[\sigma]\,\ket{\Omega^{}_{0}}}.
\eeq
Taking advantage of the arbitrariness of the initial state, we choose for $\ket{\Omega^{}_{0}}$ a Slater determinant for each fermion species constructed from single-particle plane-wave states $\phi^{}_{j}$ for $1\le j \le N/2$ with $N/2 = N^{}_{\downarrow} = N^{}_{\uparrow}$.  With this assumption, we find that the probability takes the form
\beq
P^{}_{\beta}[\sigma] = {\det}^2 \bd U^{}_{\beta}[\sigma],
\eeq
with
\beq
[\bd U^{}_{\beta}[\sigma]]_{kk'} = \bra{\phi^{}_{k}}\,\hat{\mathcal U}^{}_{\beta}[\sigma]\,\ket{\phi^{}_{k'}},
\eeq
where the indices $k,k'$ satisfy $1\le k,k' \le N/2$.

\subsubsection{Path integral form of the reduced density matrix, replica fields, and the R\'enyi entropy}

It was shown by Grover in Ref.~\cite{Grover} that the reduced density matrix can be written in terms of the fermionic creation 
and annihilation operators $\hat{c}^{\dagger}_{}, \hat{c}^{}_{}$ as a weighted average with respect to the probability measure 
$P^{}_{\beta}[\sigma]$ derived above. Specifically,
\beq
\label{Eq:rhoAGrover}
\hat{\rho}^{}_{A,\beta} = \int \mathcal D \sigma\; P^{}_{\beta}[\sigma]\;\hat{\rho}^{}_{A,\beta}[\sigma],
\eeq
where
\bea
\hat{\rho}^{}_{A,\beta}[\sigma] &=&
\det\left(\openone - G^{}_{A,\beta}[\sigma]\right) \nonumber  \times \\ && \!\!\!\!\!\!\!\!\!\!\!\! \exp\left(-\sum_{i,j \in A}\hat{c}^{\dagger}_{i}\left[\log\left(G^{-1}_{A,\beta}[\sigma]-\openone\right)\right]^{}_{ij}\hat{c}^{}_{j}\right).
\eea
It is important to note that $\hat{\rho}^{}_{A,\beta}[\sigma]$ is the reduced density matrix of a system of non-interacting fermions in the
external field $\sigma$. Expressions for non-interacting reduced density matrices were first derived in Refs.~\cite{Peschel1,Peschel2,Henley}, but 
it was not until the much more recent work of Ref.~\cite{Grover} that those were combined into the non-perturbative form of Eq.~(\ref{Eq:rhoAGrover})
amenable to Monte Carlo calculations.

In the above, $G^{}_{A,\beta}[\sigma]$ is the spatial restriction of the (equal-time) one-body density matrix for either flavor to the region $A$ computed as
\beq
G^{}_{A,\beta}[\sigma]_{rr'} = \sum_{a,b = 1}^{N/2}[\bd U^{-1}_{\beta}[\sigma]]_{ab} \;\phi^{*}_{b}(r,\beta/2)\,\phi^{}_{a}(r',\beta/2), 
\eeq
where
\bea
\phi^{}_{a}(r',\beta/2) &=& \bra{r'}\hat{\mathcal U}^{}_{\beta}[\sigma]\ket{\phi^{}_{a}}\\
\phi^{*}_{b}(r,\beta/2) &=& \bra{\phi^{}_{b}}\hat{\mathcal U}^{}_{\beta}[\sigma]\ket{r}.
\eea
We suppress the imaginary-time $\beta$ dependence in much of what follows with the understanding that calculations are to be performed in the limit of 
$\beta \to \infty$.

From this decomposed form of the reduced density matrix, an estimator for the $n$-th order R\'enyi entropy can be derived.
Because $n$ powers of $\hat{\rho}^{}_{A}$ are needed, an equal number of auxiliary fields will appear (the ``replica'' fields),
which we will denote collectively as $\bd \sigma$.

The final result (see Refs.~\cite{Grover, Assaad1, Assaad2, DrutPorter1, DrutPorter2}) takes the form
\beq
\label{Eq:EEPI}
\exp\left((1-n)S_{n,A}\right) = \Tr_{\mathcal H_A}\left[ \rho_A^n \right] = \frac{1}{Z^{n}_{}}\int \mathcal D\Sigma \;P[\bd \sigma]\;Q[\bd \sigma],
\eeq
where (note the suppressed $\beta$ dependence)
\beq
P[\bd \sigma] = P[\sigma^{}_{1}]P[\sigma^{}_{2}]\dots P[\sigma^{}_{n}],
\eeq
with the observable being
\beq
\label{Eq:EEQ}
Q[\bd \sigma] = {\det}^2 W[\bd \sigma],
\eeq
with
\bea
\label{Eq:EEW}
W[\bd \sigma]=\prod_{j=1}^{n}(\openone - G^{}_{A}[\sigma^{}_{j}])
\left[\openone + \prod_{k=1}^{n}\frac{G^{}_{A}[\sigma^{}_{k}]}{\openone - G^{}_{A}[\sigma^{}_{k}]}\right].
\eea
We have adopted a notation such that, for functions or integrals of functions of multiple auxiliary fields, we write
\beq
F[\bd \sigma] = F[\sigma^{}_1,\sigma^{}_{2},\dots,\sigma^{}_{n}],
\eeq
and
\beq
\int \mathcal D \Sigma \;F[\bd \sigma] = \int  \mathcal D\sigma^{}_1 \mathcal D \sigma^{}_{2}\dots \mathcal D\sigma^{}_{n}\;F[\bd \sigma],
\eeq
respectively.


Equation~(\ref{Eq:EEW}) poses the challenging task of inverting $\openone - G^{}_{A}$,
which can be very nearly singular, as pointed out in Ref.~\cite{Assaad1}. 
For $n\!=\!2$, no inversion is required, because the equations simplify such that
\beq
Q[\bd \sigma] = {\det}^2 
\left[(\openone - G^{}_{A}[\sigma^{}_{1}])(\openone - G^{}_{A}[\sigma^{}_{2}]) + 
{G^{}_{A}[\sigma^{}_{1}]}{G^{}_{A}[\sigma^{}_{2}]}\right].
\eeq
However, for higher $n$ there is no simplification of that kind and therefore it is less clear how 
one may avoid the problem. We solved this problem in Ref.~\cite{DrutPorter2} (see also~\cite{Humeniuk,Broecker,WangTroyer,Luitz}); 
the main point is realizing that
\beq
\label{Eq:detW}
\det \,W[\bd \sigma ] = \det \,L[\bd \sigma]\; \det \,K [\bd \sigma ],
\eeq
where
$L[\bd \sigma]$ is a block diagonal matrix (one block per replica $k$):
\beq
\label{MatrixL}
L[\bd \sigma ] \equiv \text{diag}\left[\openone-G^{}_{A}[\sigma^{}_k]\right],
\eeq
and
\beq
\label{MatrixK}
K[\bd \sigma] 
\equiv 
\left( \begin{array}{ccccccc}
\openone & 0 & 0 & \dots & 0 & \!\!\!-R[\sigma_n]\\
R[\sigma_1] & \openone & 0 & \dots & \vdots & \!\!\! 0 \\
0 & R[\sigma_2] & \openone & 0 & 0 & \!\!\! 0 \\
\vdots & \ddots & \ddots & \ddots & \openone & \!\!\! \vdots \\
0 & \dots & \dots & 0 & R[\sigma_{n-1}]  & \!\!\! \openone
\end{array} \right),
\eeq
where
\beq
R[\sigma_k] = \frac{G^{}_{A}[\sigma^{}_k]}{G^{}_{A}[\sigma^{}_k] - \openone}.
\eeq
Within the determinant of Eq.~(\ref{Eq:detW}), we multiply $K[\bd \sigma]$ and $L[\bd \sigma]$ and
define
\beq
\label{Eq:MatrixT}
T[\bd \sigma] \equiv K[\bd \sigma] L[\bd \sigma] = \openone - D\; \mathcal G[\bd \sigma],
\eeq
where $\mathcal G[\bd \sigma]$ is a block diagonal matrix defined by
\beq
\mathcal G[\bd \sigma] = \text{diag}\left[{G^{}_{A}[\sigma^{}_{k}] }\right],
\eeq
and
\beq
\label{Eq:MatrixB}
D \equiv 
\left( \begin{array}{ccccccc}
\openone& 0 & 0 & \dots & -\openone\\
\openone & \openone & 0 & \dots & 0 \\
0 & \openone & \openone  & \dots & 0 \\
\vdots & \ddots & \ddots & \ddots & \vdots \\
0 & \dots & 0 & \openone & \openone
\end{array} \right).
\eeq
Equation~(\ref{Eq:MatrixT}) is the result that allows us to bypass the inversion of $\openone - G^{}_{A}$.
Moreover, the form of $T[\bd \sigma]$ is clearly simpler than that of $W[\bd \sigma]$.
For those reasons we use $T[\bd \sigma]$ in all of the many-body calculations presented here. 
This formulation allowed us to study R\'enyi entropies as high as $n=5$; higher are also possible.

For completeness, we present here the simplification for the bosonic case as well (and add a subindex $B$ accordingly), for which 
\beq
\label{Eq:EEQB}
Q_B[\bd \sigma] = {\det}^{-2} W_B[\bd \sigma],
\eeq
and
\bea
\label{Eq:EEWB}
W_B[\bd \sigma]=\prod_{j=1}^{n}(\openone + G^{}_{A}[\sigma^{}_{j}])^{}
\left[\openone - \prod_{k=1}^{n}\frac{G^{}_{A}[\sigma^{}_{k}]}{\openone + G^{}_{A}[\sigma^{}_{k}]}\right]^{}.
\eea
The analogous strategy to avoid inversion leads here to 
\beq
\label{Eq:MatrixT}
T_B[\bd \sigma] \equiv \openone - D_B \mathcal G[\bd \sigma],
\eeq
where $\mathcal G[\bd \sigma]$ is a block diagonal matrix defined by
\beq
\mathcal G[\bd \sigma] = \text{diag}\left[{G^{}_{A}[\sigma^{}_{n}] }\right],
\eeq
and
\beq
\label{Eq:MatrixB}
D_B \equiv 
\left( \begin{array}{ccccccc}
-\openone& 0 & 0 & \dots & \openone\\
\openone & -\openone & 0 & \dots & 0 \\
0 & \openone & -\openone  & \dots & 0 \\
\vdots & \ddots & \ddots & \ddots & \vdots \\
0 & \dots & 0 & \openone & -\openone
\end{array} \right).
\eeq

\subsubsection{Signal-to-noise issues and how to overcome them}

The path integral form of the R\'enyi entropy Eq.~(\ref{Eq:EEPI}) has a deceptively simple form:
It seems obvious that one should interpret $P[\bd \sigma]$ as the probability density and $Q[\bd \sigma]$
as the observable being averaged. This, in some sense, is a trap: while $Q[\bd \sigma]$ is crucially
sensitive to correlations among the replica fields $\sigma_k$, $P[\bd \sigma]$ completely factorizes across replicas
(i.e. it is insensitive to said correlations). As a consequence, a Monte Carlo implementation sampling $\bd \sigma$ 
according to $P[\bd \sigma]$ will give outlandish values of $Q[\bd \sigma]$ that fluctuate wildly and may not 
converge to the expected value. This feature is what in the lattice QCD area is often called an 
overlap problem (see e.g. Refs~\cite{NoiseSignProblemStatistics,LogNormalDeGrand}). The present case is especially 
challenging in 2D and 3D, as the magnitude of $Q[\bd \sigma]$ is expected to grow exponentially with the size of the 
boundary of the subregion $A$ (see e.g.~\cite{Humeniuk, Broecker}).

Motivated by the similarity between the numerator of Eq.~(\ref{Eq:EEPI}) and the conventional path-integral form of
partition functions, we address the overlap problem by first differentiating with respect to a parameter, then
using Monte Carlo methods to compute that derivative, and finally integrating at the end. We outline this
procedure in detail in Ref.~\cite{DrutPorter2}, and reproduce part of it here.

\begin{figure}[t]
\includegraphics[width=1.0\columnwidth]{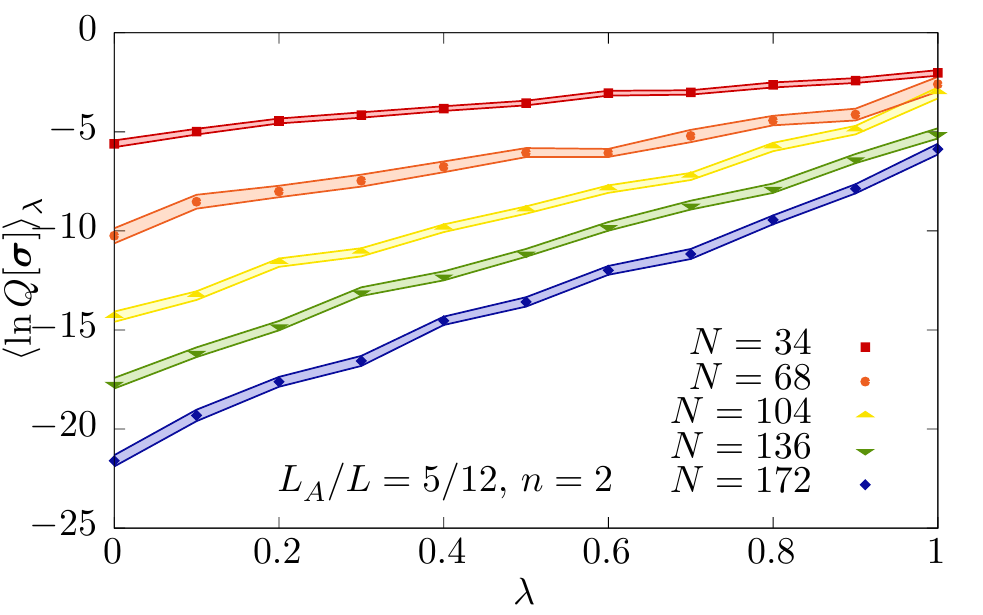}
\caption{\label{Fig:EEIRegionPart}$\lambda$ dependence of $\langle\ln{Q}[{\bd \sigma}]\rangle_\lambda^{}$ for a subsystem 
of size $L_A=5/12L$,
for $N=34,68,104,136,172$ fermions at unitarity in 
a box of size $L=N_x\ell$ (where $N_x=12$ points and $\ell=1$), and for R\'enyi order $n=2$.
Similar plots are obtained by varying, instead of the particle number, the region size
and the R\'enyi order. These are shown in Appendix~\ref{Appendix2}.
}
\end{figure}

We introduce a parameter $0\le\lambda \le 1$ by defining a function $\Gamma(\lambda;g)$ such that
\beq
\label{Eq:GammaDef}
\Gamma(\lambda;g) \equiv \int \mathcal D \Sigma \; P[{\bd \sigma}]\;
Q^{\lambda}[{\bd \sigma}].
\eeq
Normalization of $P[{\bd \sigma}]$ implies that
\beq
\ln \Gamma(0;g)=0,
\eeq
while Eq.~(\ref{Eq:EEPI}) implies
\beq
\ln \Gamma(1;g)=(1-n)S_{n,A}^{}.
\eeq
Using Eq.~(\ref{Eq:GammaDef}), 
\beq
\label{Eq:dlnGammadlambda}
\frac{\partial \ln \Gamma}{\partial \lambda}=
\int \mathcal D \Sigma \; \tilde P[{\bd \sigma};\lambda]\; \ln Q[{\bd \sigma}],
\eeq
where
\beq
\label{Eq:Ptilde}
\tilde{P}[{\bd \sigma};\lambda]\equiv\frac{1}{\Gamma(\lambda;g)}P[{\bd \sigma}]\;Q^{\lambda}[{\bd \sigma}]
\eeq
is a well-defined, normalized probability measure which 
features the usual weight $P[{\bd \sigma}]$ as well as an entanglement contribution $Q^{\lambda}[{\bd \sigma}]$. 
It is the latter factor that induces entanglement-specific correlations in the sampling of $\bd \sigma$ when
probability $\tilde{P}[{\bd \sigma};\lambda]$. 

\begin{figure}[t]
\includegraphics[width=1.0\columnwidth]{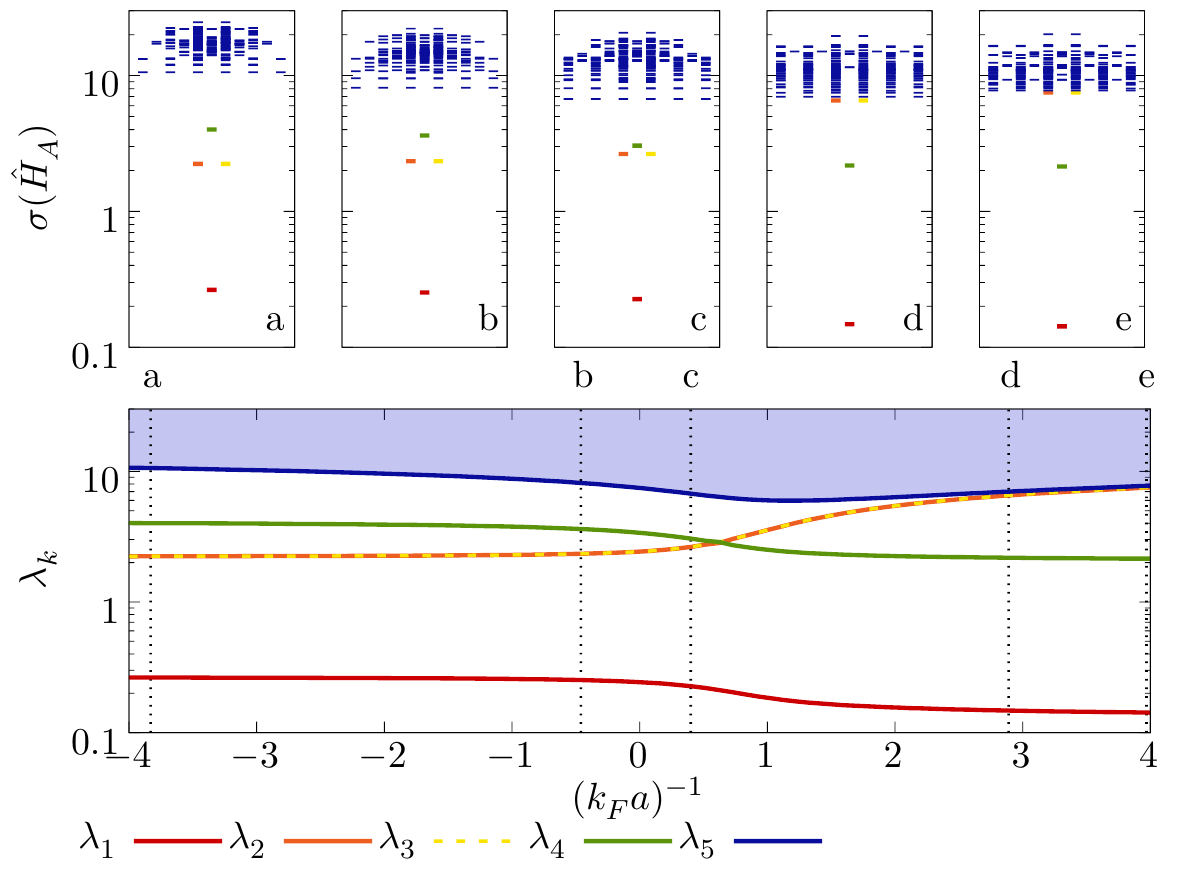}
\caption{\label{Fig:EESpectrum} Bottom panel: Low-lying entanglement spectrum of the two-body problem as a function
of the dimensionless coupling $(k_F a)^{-1}$ in the BCS-BEC crossover, for a cubic subregion $A$ of linear size $L_A/L = 0.5$.
Top panels (a - e): Low-lying (and part of the high) entanglement spectrum for selected couplings (a - e) at the top of the bottom
panel.}
\end{figure}

Thus, $S^{}_{n,A}$ is calculated by using $\lambda=0$ as a reference point and computing $S_{n,A}$ via
\beq
\label{Eq:SnMCFinal}
S_{n,A}^{} = \frac{1}{1-n}
\int_{0}^{1}d\lambda\;\langle\ln{Q}[{\bd \sigma}]\rangle_\lambda^{},
\eeq
where
\beq
\langle X\rangle_\lambda^{}=\int \mathcal D \Sigma \; \tilde{P}[{\bd \sigma};\lambda]\; X[{\bd \sigma}].
\eeq
We thus obtain an integral form of the interacting R\'enyi entropy
that can be computed using any MC method (see e.g.~\cite{MCReviews2, MCReviews3, MCReviews4}), in particular hybrid Monte Carlo~\cite{HMC1, HMC2}
to tackle the evaluation of $\langle\ln{Q}[{\bd \sigma}]\rangle_\lambda^{}$ as a function of $\lambda$.
In practice, we find that $\langle\ln{Q}[{\bd \sigma}]\rangle_\lambda^{}$ is a smooth function of $\lambda$,
as exemplified in Fig.~\ref{Fig:EEIRegionPart}.
It is therefore sufficient to perform the numerical integration using a uniform grid.


\section{Results: Two-body system \label{sec:Results1}}

We solve the two-body problem via the projection method outlined previously, which furnishes the
full two-body wavefunction on the lattice. We ensure that the continuum limit is approached by solving
the problem for multiple lattice sizes, and by computing the renormalized coupling using the energy spectrum and 
L\"uscher's formalism~\cite{Luescher1, Luescher2}. The latter indicates that the relationship between the energy eigenvalues and the
scattering phase shift $\delta(p)$ is given by
\beq
\label{Luscherformula}
p \cot \delta(p) = \frac{1}{\pi L}\mathcal S^{}(\eta)
\eeq
where $\eta = \frac{pL}{2\pi}$ and $L$ is the box size, such that the energy of the two-body problem is $E = p^2/m$; and
\beq
\label{Seta}
\mathcal S^{}(\eta) \equiv \lim_{\Lambda \to \infty} \left ( \sum_{\bf n}^{} 
\frac{\Theta(\Lambda^2 - {\bf n}^2)}{{\bf n}^2 - \eta^2} - 4 \pi \Lambda \right ),
\eeq
where the sum is over all 3D integer vectors, and $\Theta(x)$ is the Heaviside function. 
In turn, the scattering phase shift determines the scattering parameters via
\beq
\label{EffectiveRangeExpansion}
p \cot \delta(p) = -\frac{1}{a}  + \frac{1}{2} r^{}_\text{eff} p^2 + O(p^4),
\eeq
where $\delta$ is the scattering phase shift, $a$ is the scattering length, and $r^{}_\text{eff}$ is the effective range.

\subsection{Low-lying entanglement spectrum}

Once the matrix elements of $\hat \rho_A$ are calculated from the projected ground state, as shown above,
we obtain the eigenvalues using standard diagonalization routines to obtain the entanglement spectrum $\sigma(\hat H_A)$, 
which is defined as the spectrum of the entanglement Hamiltonian $\hat H_A$, where
\beq
\hat \rho_A = e^{-\hat H_A}.
\eeq

\begin{figure}[t]
\includegraphics[width=1.0\columnwidth]{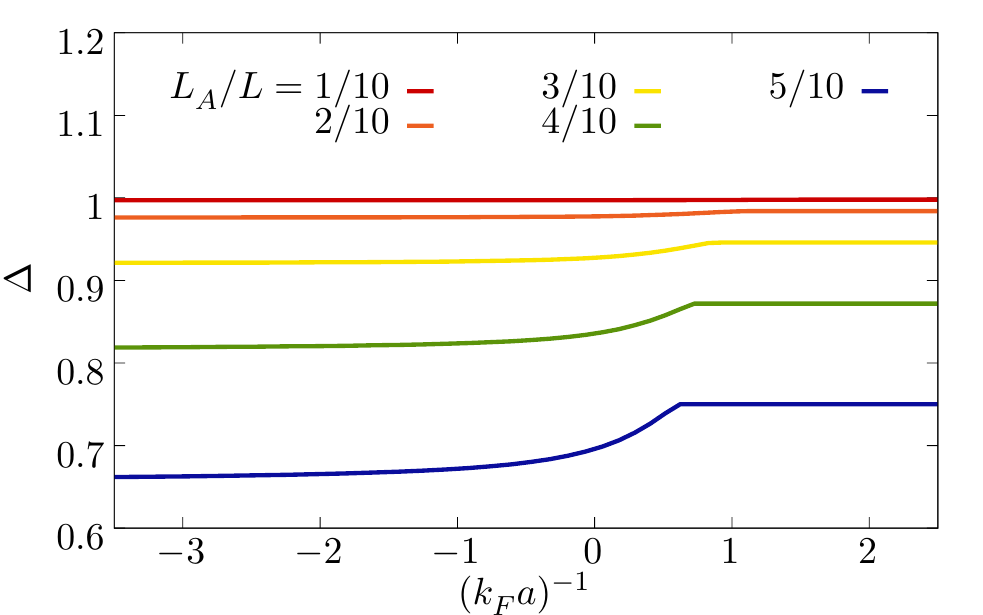}
\caption{\label{Fig:SchmidtGap} Schmidt gap $\Delta$ between the two largest eigenvalues of the reduced density matrix,
at $L_A/L=0.1,0.2,...,0.5$ (top to bottom), for the two-body system as a function of the coupling $(k_Fa)^{-1}$}
\end{figure}

In Fig.~\ref{Fig:EESpectrum}, we present our results for $\sigma(\hat H_A)$ for a cubic subregion $A$ of linear size $L_A/L = 0.5$, 
for two particles in the BCS-BEC crossover, parametrized by the dimensionless coupling $(k_F a)^{-1}$, where $k_F$ is the Fermi 
momentum (merely a measure of the particle density in the periodic box, as for two particles there is of course no Fermi surface) 
and $a$ is the s-wave scattering length. The latter was determined using the L\"uscher formalism outlined above.

The main features of $\sigma(\hat H_A)$ can be described as follows. We note first that beyond the lowest 4 or 5 eigenvalues, shown
as $\lambda_1$ to $\lambda_5$ in the bottom panel of Fig.~\ref{Fig:EESpectrum}, the multiplicity of eigenvalues grows dramatically, 
forming a quasi-continuum. For this reason, we focus here on the lowest 5 eigenvalues and characterize the rest statistically in the next 
section. As is evident from the figure, the dependence of all $\lambda_k$ on $(k_F a)^{-1}$ is rather mild and smooth, although it has at 
a few crisp features: there is a rather large gap between $\lambda_1$ and the next eigenvalue, which implies that the R\'enyi entanglement entropies
are dominated by that eigenvalue; there is a crossing of $\lambda_2$, $\lambda_3$ 
and $\lambda_4$ on the BEC side of the resonance; after that crossing $\lambda_2$ and $\lambda_3$ heal to $\lambda_5$ and effectively merge 
into the lower edge of the quasi-continuum part of the spectrum. The evolution of these properties along the crossover is shown in detail in panels 
a -- e of Fig.~\ref{Fig:EESpectrum}.

In Fig.~\ref{Fig:SchmidtGap} we show the Schmidt gap $\Delta$ (see Refs~\cite{DeChiara}), defined as the separation between the two largest eigenvalues
of the reduced density matrix $\hat \rho_A$, for $L_A/L=0.1,0.2,...,0.5$, as a function of $(k_Fa)^{-1}$. Since we do not expect a quantum phase transition as 
a function of $(k_Fa)^{-1}$, we similarly do not expect the Schmidt gap to vanish. As a result of the eigenvalue crossing explained above, however,
there exists a sharp change (in the sense of a discontinuous derivative) in $\Delta$ in the BCS-BEC crossover, which takes place in the strongly coupled
region $0 < (k_Fa)^{-1} < 1$. It is also evident that, because $\lambda_1$ and $\lambda_4$ track each other at a very nearly constant separation, the 
Schmidt gap becomes constant to the right of the sharp edge in Fig.~\ref{Fig:SchmidtGap}. As with other features of this spectrum, it remains to 
be determined how $\Delta$ evolves as a function of particle number, in particular as a Fermi surface forms and Cooper pairing correlations emerge.

\begin{figure}[b]
\includegraphics[width=1.0\columnwidth]{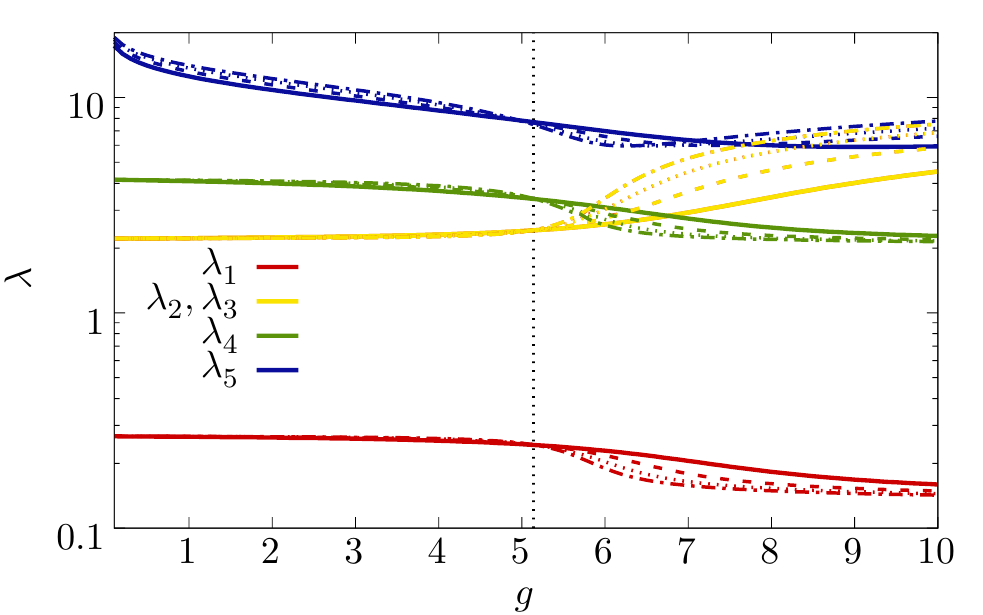}
\caption{\label{Fig:EESpectrumSizeEffects} Entanglement spectrum of the two-body problem in the BCS-BEC crossover 
as a function of the bare lattice coupling at different lattice sizes: solid, dashed, dotted, dash-dotted, for $N_x = 4, 6, 8, 10$, respectively. 
The subsystem size was fixed to $L_A/L = 0.5$.
The coupling corresponding to the unitary point is marked with a vertical dashed line. Note how different volumes cross precisely at unitarity, which reflects
the property of scale invariance.}
\end{figure}

As mentioned above, our calculations were carried out in a periodic box. We show the corresponding size effects in 
Fig.~\ref{Fig:EESpectrumSizeEffects}, where we show the entanglement spectrum of the two-body system as a function of the 
bare lattice coupling $g$. In that figure, it is clear that finite-size effects are smallest on the BCS side of the resonance,
but become considerably more important on the BEC side. This is consistent with the expectation that, once a two-body bound state
forms (as the coupling is increased away from the non-interacting point), the sensitivity to lattice-spacing effects is enhanced. It is noteworthy,
in particular, that one may identify the unitary regime just by looking at this figure: for any given eigenvalue, the data for different lattice sizes 
crosses at about the same value of $g$; this is reminiscent of the finite-size scaling behavior of order parameters in critical phenomena, as it is 
the hallmark of scale invariance at phase transitions.

The process of reducing finite-size effects, at fixed particle number, implies approaching the dilute limit, i.e. using larger lattices.
When that limit is approached, the renormalization prescription that replaces $g$ with the physical coupling $(k_F a)^{-1}$ (described above) 
should force the finite-size calculations to collapse to a single, universal (in the sense of size-independent) curve. This is indeed what we find 
and what yields the results of Fig.~\ref{Fig:EESpectrum}.

\subsection{High entanglement spectrum}

\begin{figure}[t]
\includegraphics[width=1.00\columnwidth]{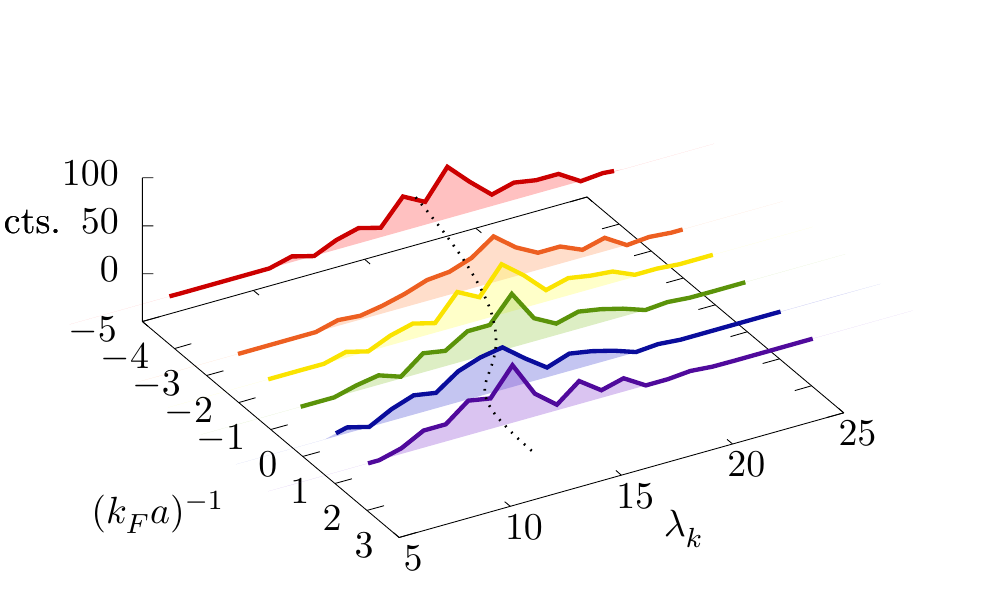}
\includegraphics[width=0.97\columnwidth]{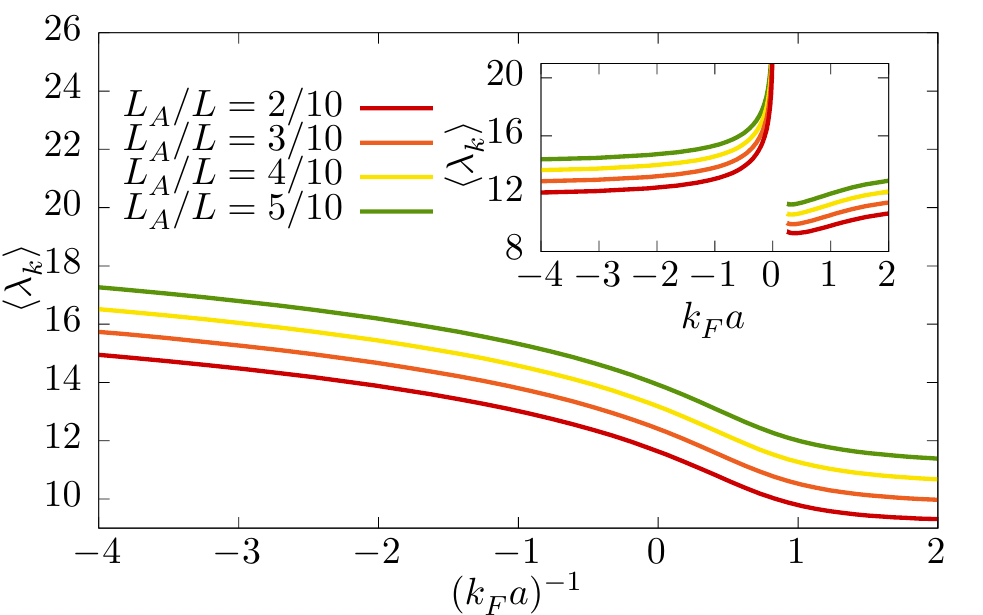}
\includegraphics[width=0.97\columnwidth]{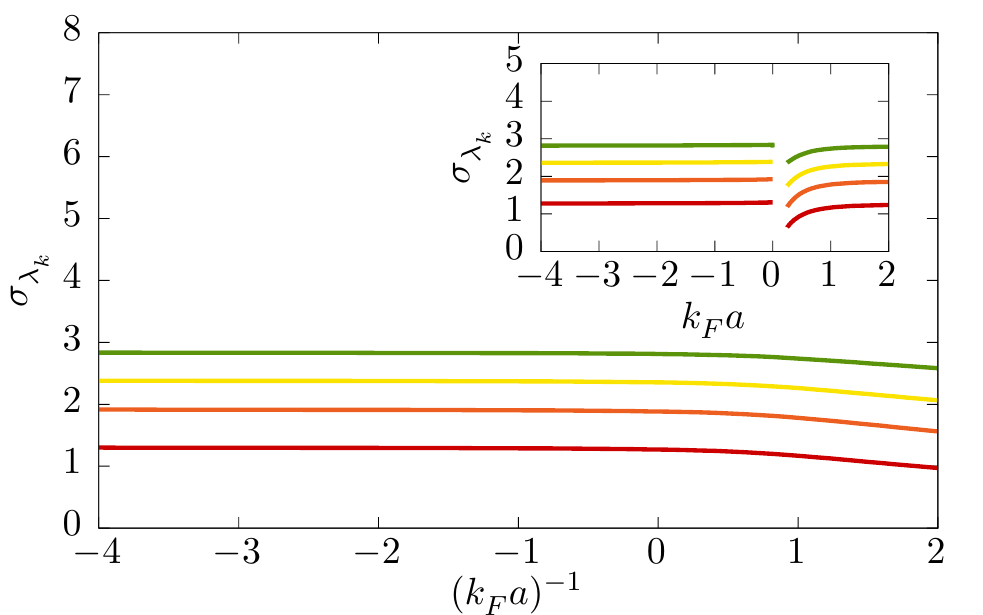}
\caption{\label{Fig:HighEESpectrum} Top: Histogram of the high entanglement spectrum of the two-body problem showing
number of counts (cts.) as a function of the coupling $(k_F a)^{-1}$ and the entanglement eigenvalue $\lambda_k$,
for region size $L_A / L = 1/2$. The dashed line shows the dependence of the mean (see also middle plot).
Middle and bottom: Mean and standard deviation, respectively, of the high entanglement spectrum distribution, as a functions 
of the interaction strength $(k_F a)^{-1}$ (main) and $k_F a$ (inset). In each plot the different curves show
results for various $L_A/L$. Note that the weak coupling limit corresponds to $k_F a \to 0^-$.
}
\end{figure}

As mentioned in the previous section, the entanglement spectrum $\sigma(\hat H_A)$ above $\lambda_5$, which we will refer to here 
as the high entanglement spectrum, displays a rapidly growing multiplicity of eigenvalues which we deem best to analyze using elementary 
statistical methods. In Fig.~\ref{Fig:HighEESpectrum}, we show the eigenvalue distribution of the high entanglement spectrum for different 
system sizes, in histogram form. More importantly, we find that the mean and standard deviation of that distribution, shown here in 
Fig.~\ref{Fig:HighEESpectrum} (middle and bottom), are smooth functions of $(k_F a)^{-1}$; the mean, in particular, diverges as the coupling is
turned off. We interpret this effect as strong evidence that the high sector of $\sigma(\hat H_A)$ is a non-perturbative component of $\hat H_A$ 
that is entirely due to quantum fluctuations induced by the interaction. Although the two-body system has no Fermi surface, it seems natural to
conjecture a link between Cooper pairing and the high entanglement spectrum. Determining whether this is true, however, is a challenging
problem that requires studying the high entanglement spectrum in the progression from few to many particles.

Our numerical calculations show a large number of eigenvalues that lie far (at least 9 to 10 orders of magnitude) above the
high entanglement spectrum. While we cannot discard that those eigenvalues are consistent with numerical
noise (they come from the lowest eigenvalues of the reduced density matrix), there are enough of them to warrant this brief comment.
Although there is a large number of such eigenvalues, their contribution to the entanglement entropy is considerably suppressed by their 
small magnitude. We add to this discussion below.

\subsection{Entanglement entropy}

Using our knowledge of the eigenvalues $\lambda_k \in \sigma(\hat H_A)$, the entanglement entropy of the two-body problem is easily determined. 
Indeed, the von Neumann entropy is
\beq
\label{Eq:SvN}
S^{}_{\mbox{vN},A} = -\Tr_{\mathcal{H}^{}_{A}}\left [ \hat{\rho}^{}_A\ln\hat{\rho}^{}_A\right ] = \sum_{k}^{} \lambda_k \; e^{-\lambda_k},
\eeq
and the $n$-th order R\'enyi entanglement entropy is
\beq
\label{Eq:SRn}
S^{}_{n,A} = \frac{1}{1-n}\ln \Tr_{\mathcal{H}^{}_{A}} \left[ \hat{\rho}^{n}_{A}  \right]= \frac{1}{1-n} \ln \sum_{k}^{} e^{-n\lambda_k}.
\eeq

\begin{figure}[t]
\includegraphics[width=1.0\columnwidth]{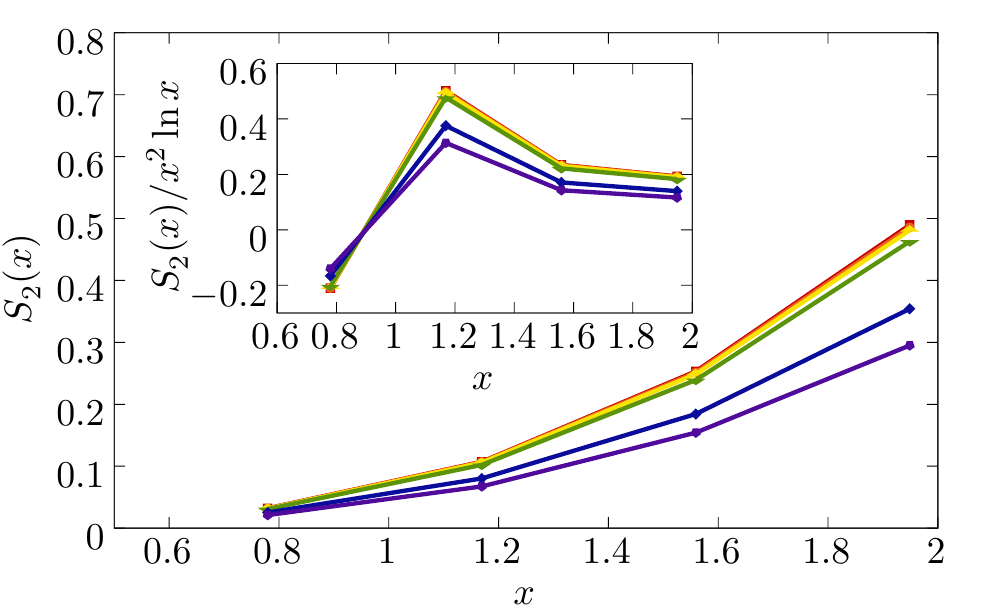}
\includegraphics[width=1.0\columnwidth]{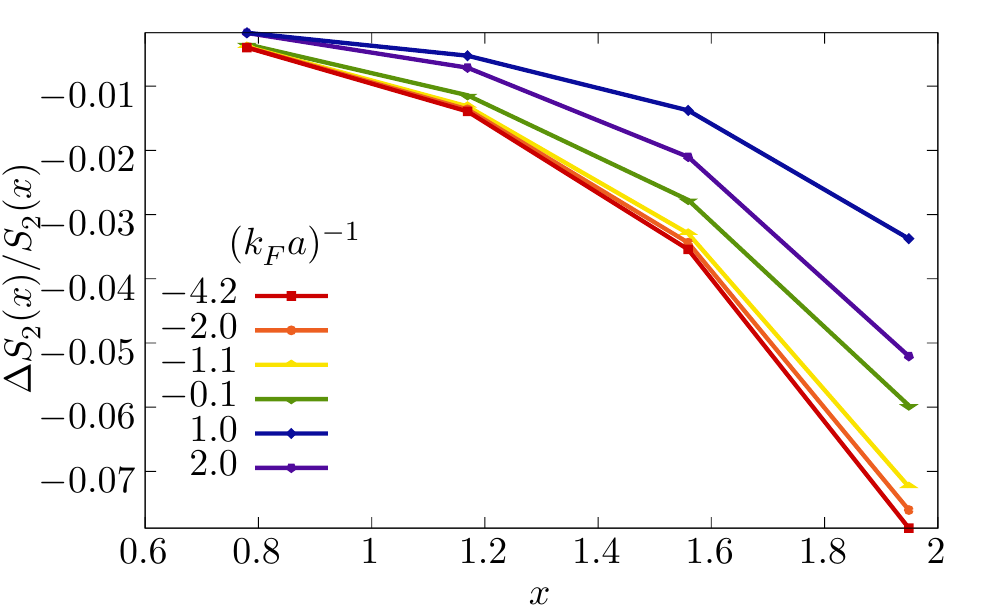}
\caption{\label{Fig:DeltaHighEESpectrum}
Top: Second R\'enyi entanglement entropy $S_2$ of the two-body problem as a function of $x=k_F L_A$ and 
for several values of the coupling $(k_Fa)^{-1}$. Inset: $S_2$ scaled by $x^2 \ln x$.
Bottom: Relative contribution of the high entanglement spectrum to the second R\'enyi 
entanglement entropy $S_2$, as a function of $x=k_FL_A$.
}
\end{figure}

In Fig.~\ref{Fig:DeltaHighEESpectrum} (top panel), we show $S^{}_{2}$ as a function of $x = k_FL_A$ and the coupling $(k_F a)^{-1}$.
Remarkably, the trend towards the leading asymptotic behavior proportional to $x^2 \ln x$ appears to set in
at $x \simeq 2$ for all couplings. This is surprising, as there is no obvious reason for this to be the case. 
As we will see below, we find the same kind of behavior for the many-body Fermi gas at resonance. 

To show explicitly the effect of the high entanglement spectrum on $S_2$, which we referred to in the previous section, we show 
in Fig.~\ref{Fig:DeltaHighEESpectrum} (bottom panel) the contribution $\Delta S_2$ of the first entanglement eigenvalue to the full $S_2$. It is clear in that plot 
that the contribution is at most on the order of $8\%$ for the parameter ranges we studied.

\section{Results: Many-body system \label{sec:Results2}}

Using the many-body lattice Monte Carlo techniques described above, along with the tuning procedure
outlined in the previous section, we computed several entanglement entropies of the unitary Fermi gas,
aiming to characterize its leading and sub-leading asymptotic behavior as a function of the subregion
size $x = k_F L_A$.

The results shown throughout this section were obtained by gathering 250 decorrelated auxiliary field configurations (where a 
single ``auxiliary field'' contains all the replicas required to determine the desired R\'enyi entropy) for each value of the auxiliary 
parameter $\lambda$.
We used particle numbers in the range $N= 4 - 400$ and cubic lattice sizes in the range $N^{}_x = 6 - 16$ with
periodic boundary conditions. The projection to the ground state was carried out by extrapolation to the limit of
large imaginary-time direction. The auxiliary parameter $\lambda$ was discretized using $N_\lambda = 10$ points,
which we found to be enough to capture the very mild dependence on that parameter, as explained in a previous section
(see also Appendix~\ref{Appendix2} for further details).

Because the methods we implemented impose a discretization of spacetime, special attention was given to 
the ordering of the scales, to ensure that the thermodynamic and continuum limit were approached.
Specifically, we required the following ordering:
\beq
\label{Eq:Windows}
k_F\ell \ll 1  \ll k_F L_A \ll k_F L,
\eeq
where $\ell= 1$ is the lattice spacing, $L_A$ is the subsystem size, and
$L = N_x \ell = N_x$ is the full system size. The first condition on the left of Eq.~(\ref{Eq:Windows}) ensures that the continuum limit 
is approached; the second condition implies that the region determined by $L_A$ must contain many particles (since the density is the 
only scale in the system, this condition defines the large-$L_A$ regime); and the last condition means that $L_A \ll L$, to ensure finite-size 
effects are minimized.
This ordering was accomplished by carefully choosing the restrictions on $L_A$ for a given particle number $N$, while 
aiming to maintain a large $N$. The latter, however, requires $L$ to be large in order to avoid high densities where $k_F \simeq 1$, 
which can be sensitive to lattice-spacing effects. In addition, we set $L_A \leq 0.45 L$ as a compromise to satisfy the last inequality.

\begin{figure}[t]
\includegraphics[width=1.0\columnwidth]{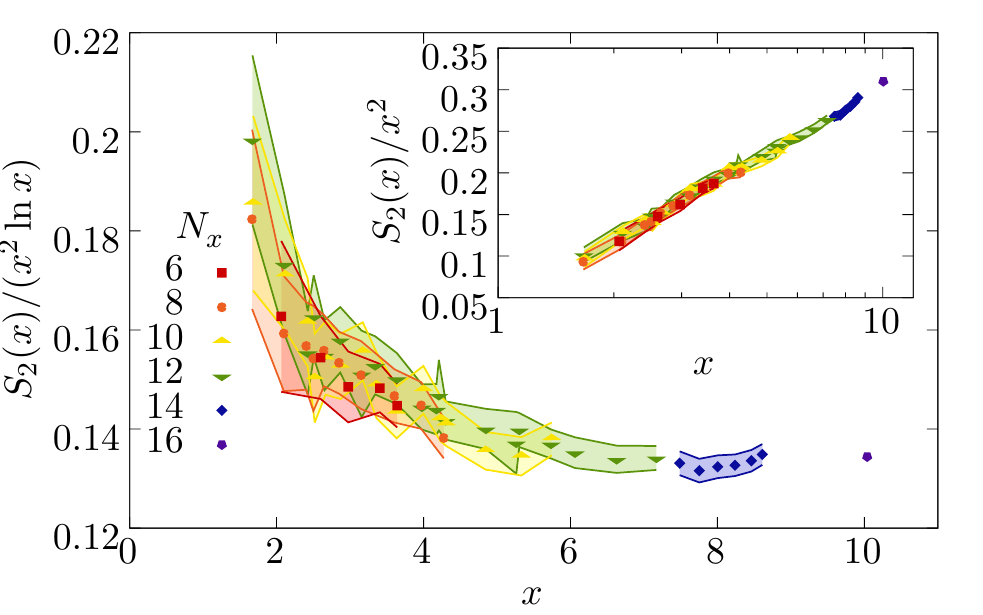}
\caption{\label{Fig:ImpalaVolOrderTwo} Second R\'enyi entropy of the unitary Fermi gas in units of $x^2 \ln x$ (main)
and $x^2$ (inset), where $x=k_FL_A$. Note the linear scale in the main plot and logarithmic scale in the inset. Although the range of values of $x$
is limited by our computational power (as set by method and hardware), the fact that the main plot is consistent with a straight line
is a strong indication that the leading behavior of the entanglement entropy as a function of $x$ is logarithmic. Moreover, we see that
that behavior sets in as early as $x \simeq 2$, which is roughly consistent with the non-interacting case shown in Fig.~\ref{Fig:scalingFG}.
}
\end{figure}

In Fig.~\ref{Fig:ImpalaVolOrderTwo} we show our results of the second R\'enyi entropy $S_2$ of the unitary Fermi gas
in volumes of $N_x^3$ lattice points, where $N^{}_x = 6 - 16$, as a function of $x = k_F L_A$, for cubic subsystems
of side $L_A$. Within the statistical uncertainty, shown in colored bands,
the results for different volumes coincide, which indicates that our results are in the continuum
and thermodynamic regimes.

The inset of Fig.~\ref{Fig:ImpalaVolOrderTwo} shows $S_2$ scaled by $x^2$ in a semi-log plot.
The fact that the trend is clearly linear supports the assertion that $x$ is large enough to discern the asymptotic
regime, where $S_2/x^2 \propto \ln x$. As in the case of the non-interacting Fermi gas, mentioned in the Introduction,
this onset of the asymptotic regime appears to be at $x\simeq 2$.

\subsection{R\'enyi entanglement entropies}

\begin{figure}[t]
\includegraphics[width=1.0\columnwidth]{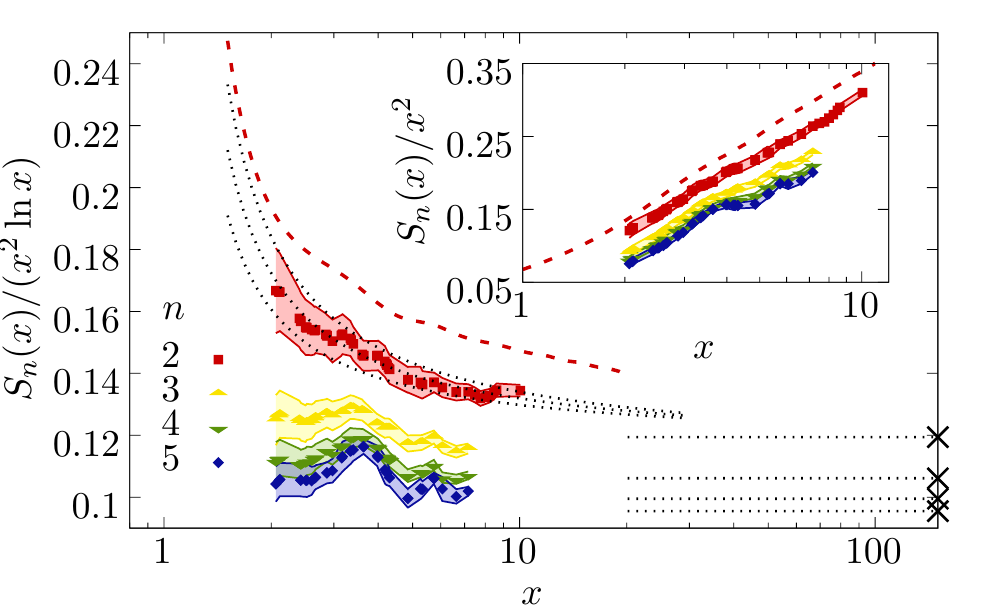}
\caption{\label{Fig:ImpalaOrders} R\'enyi entropies of order $n=2,3,4,5$ (data points with error bands in red, yellow, green, and blue, respectively) 
of the unitary Fermi gas in units of $x^2 \ln x$ (main plot) and $x^2$ (inset), where $x=k_FL_A$. Note the logarithmic
scale in the $x$ axis. The red dashed line shows the non-interacting result for $n=2$, obtained using the overlap matrix method.
The black dotted lines plotted over the $n=2$ data correspond to a fit the functional form $f(x) = a + b/\ln(x)$ (central line, with uncertainties 
marked by upper and lower dotted lines).
The crosses on the right, and the corresponding horizontal dotted lines, indicate the expected
asymptotic value $c(n)$ (from top to bottom, for $n=2,3,4,5$) for a non-interacting gas (see Refs.~\cite{NonIntS1,NonIntS2,NonIntS3,NonIntS4,NonIntS5}), 
which we reproduce in Eq.~(\ref{Eq:cn}); numerically, they are $c(2)=0.11937...$, $c(3)=0.10610...$, $c(4)=0.09947...$, and $c(5)=0.09549...$ .
}
\end{figure}

Using the formalism presented above for the determination of R\'enyi entanglement entropies
for $n\geq2$, we computed $S_{n,A}$ for the resonant Fermi gas for $n=2,3,4,5$, as a function of $x = k_F L_A$. 
In Fig.~\ref{Fig:ImpalaOrders} we show our main results. 
To interpret those results, we briefly discuss the noninteracting case. In Refs.~\cite{NonIntS1,NonIntS2,NonIntS3,NonIntS4,NonIntS5} it 
was shown that the leading-order behavior or the entanglement entropy of non-interacting 3D fermions as a function of $x$ is given by
\beq
S_{n,A}(x) = c(n) x^2 \ln x + o(x^2),
\eeq
where 
\beq
\label{Eq:cn}
c(n) = \frac{1 + n^{-1}_{}}{24(2\pi)^{d-1}_{}}\int^{}_{\partial\Omega}\int^{}_{\partial\Sigma}dS^{}_{x}dS^{}_{k}\;|{\bf \hat{n}}^{}_{x}\cdot{\bf \hat{n}}^{}_{k}|
\eeq
where $\Omega$ is the real-space region $A$ scaled to unit volume with normal ${\bf \hat{n}}^{}_{x}$, $\Sigma$ is the Fermi volume scaled by the Fermi momentum 
with unit normal ${\bf \hat{n}}^{}_{k}$. In our case, $A$ is a cubic subsystem (as in Fig.~\ref{Fig:Region}) and a spherical Fermi volume.
 
The noninteracting case is shown in Fig.~\ref{Fig:ImpalaOrders} in two ways. The asymptotic result at large $x$ is shown with crosses 
on the right edge of the plot, extended into the plot (as a visual aid) with dashed black horizontal lines for $n=2,3,4,5$ (top to bottom). 
With a thick red dashed line we show the case $n=2$ at finite $x$, as obtained with the overlap-matrix method~\cite{OverlapMatrixMethod}.

Our results for $S_{n,A}$ for the unitary Fermi gas (data points with error bands) appear to heal to the noninteracting limit when
the slow decay (see below) to a constant at large $x$ is taken into account; this statement holds especially in the $n=2$ case 
where the sub-leading oscillations allow for a relatively clean fit. Indeed, our fits for $n=2$ give
\beq
S_{2,A}(x) = a x^2 \ln x + b x^2,
\eeq
with $a= 0.114(2)$ and $b= 0.04(1)$, while Eq.~(\ref{Eq:cn}) yields $c(2) = 3/(8\pi)\simeq 0.119366\dots$.
While $c(2)$ are different to within our uncertainties, they are surprisingly close (between $3$ and $6\%$).
The sub-leading behavior is consistent with an area law $\propto x^2$.
As $n$ is increased, sub-leading oscillations become increasingly 
apparent; however, they are mild enough that it is still possible to discern the asymptotic behavior at large $x$.
For $n=3,4,5$, oscillations notwithstanding, the results in the large-$x$ limit appear again to be close to the noninteracting case.

\begin{figure}[t]
\includegraphics[width=1.0\columnwidth]{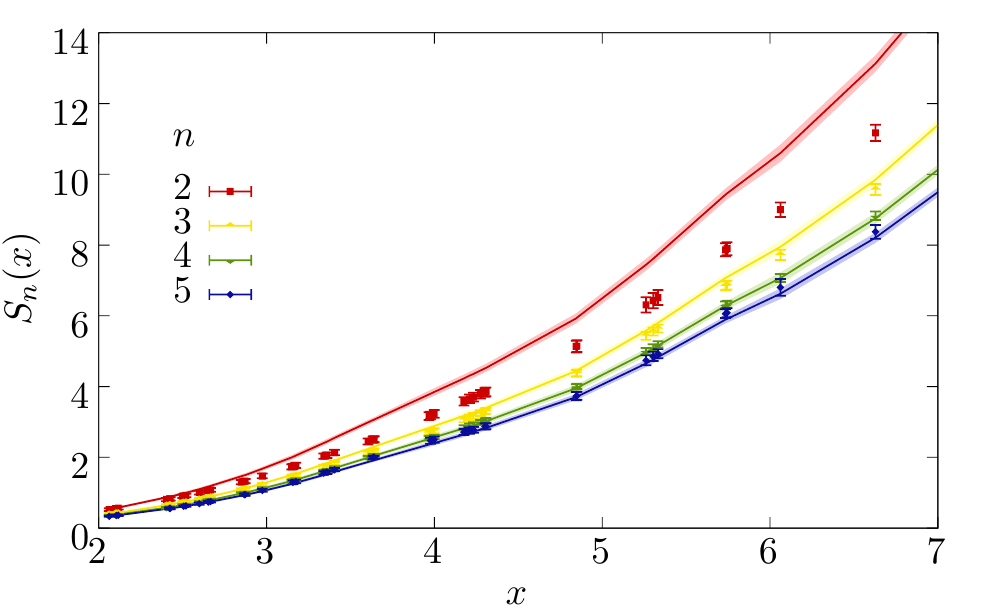}
\caption{\label{Fig:LowestEval} R\'enyi entanglement entropy $S_n$ as a function of $x = k_F L_A$ for
$n=2,3,4,5$ (top to bottom). Monte Carlo results are shown as data points with error bars. The 
solid lines show the result of computing $S_n$ using only the lowest entanglement eigenvalue $\lambda_1$, 
i.e. the approximation $S_n = \frac{n}{n-1} \lambda_1$.
Uncertainties appear as shaded regions around the central value.
}
\end{figure}

Using our results for the entanglement entropies $S_n$ as a function of $n$, it is possible to use the power method to extract the lowest 
eigenvalue $\lambda_1$ of the entanglement spectrum as a function of $x$. We studied the decay of $(1-n) S_n / n$ to a constant value which, 
given Eq.~(\ref{Eq:SRn}), we identified as $-\lambda_1$.
In Fig.~\ref{Fig:LowestEval} we show the result of using that sole eigenvalue to approximate $S_n$. As expected, higher orders $n$
emphasize the contribution from the lowest entanglement eigenvalue (highest eigenvalue of the reduced density matrix), which 
progressively dominates $S_n$ as $n$ increases. From the $n$ dependence of $S_n$, it is also possible to study the degeneracy of the lowest 
entanglement eigenstate; at large $n$,
\beq
\frac{(1-n)}{n} S_n \simeq \frac{\ln d_1}{n} - \lambda_1 + \dots,
\eeq
where the ellipsis indicates exponentially suppressed terms, and $d_1$ is the degeneracy associated with $\lambda_1$.
We find a vanishing first term, which indicates that $d_1$ is consistent with unity.

\section{~\label{sec:Conclusions}Summary and conclusions}

We implemented two different lattice methods to characterize non-perturbatively the
entanglement properties of three-dimensional spin-$1/2$ fermions in the strongly interacting, resonant regime 
of short interaction range and large scattering length, i.e. the unitary limit.
This regime is scale invariant (in fact, non-relativistic conformal invariant) in the sense that it presents as many scales as non-interacting
gases and therefore its properties are universal characteristics of three-dimensional quantum mechanics, i.e.
in the same sense as critical exponents that characterize phase transitions.

We analyzed the two-body spectrum of the entanglement Hamiltonian along the BCS-BEC crossover and presented
results for the low-lying part, which displays clear features 
as the strength of the coupling is varied, such as eigenvalue crossing close to the resonance point and 
merging in the BEC limit. The lowest two eigenvalues in the spectrum correspond to the largest two eigenvalues of the reduced density matrix, 
which are separated by the Schmidt gap. We found that the latter displays a sharp change at strong coupling, in the vicinity of the conformal point
$(k_Fa)^{-1} = 0$.

We also carried out a statistical characterization of the high entanglement spectrum, 
which appears as a quasi-continuum distribution with well defined mean and standard deviation, which we mapped out 
along the crossover. We found that the mean of the distribution tends to infinity in the noninteracting limit, 
which indicates that that sector is due to non-perturbative effects in the entanglement Hamiltonian. 
In contrast, the low-lying part of the spectrum has a finite noninteracting limit.
All of the above two-body results were obtained with non-perturbative non-stochastic
methods which are easily generalizable to higher particle numbers (as we show analytically and diagramatically for 3 particles).

In addition, we studied the R\'enyi entropies of degree $n=2,3,4$, and $5$ of many fermions in the unitary limit,
which we calculated using a method recently developed by us (based on an enhanced version of the algorithm of Ref.~\cite{Grover}). 
We found that, remarkably, the large $x=k_F L_A$ (i.e. subsystem size) limit for those entanglement entropies sets in for $x$ as low as $2.0$,
which allowed us to characterize the leading and sub-leading asymptotic behavior using $2 \leq x \leq 10$.
For entropies of order $n > 2$, on the other hand, we found that sub-leading oscillations are enhanced, but
not enough to spoil the visualization of the asymptotic behavior at large $x$.

Our experience with Monte Carlo calculations of $S_{n,A}$ in 1D gave us empirical indication that
the entanglement properties of the unitary Fermi gas might not be too different from those of a non-interacting gas. 
However, since unitarity corresponds to a strongly correlated, three-dimensional point, that intuition could very well 
have been wrong. Our calculations indicate that the leading-order asymptotic behavior is approximately consistent with 
that of a non-interacting system, while the sub-leading behavior is clearly different.

The recent measurement of the second R\'enyi entropy of a bosonic gas in an optical lattice~\cite{Greiner1,Greiner2} shows that
it is possible to experimentally characterize the entanglement properties of the kind of system analyzed here.
Our calculations are therefore predictions for such experiments for the case of fermions tuned to the unitary limit.

\acknowledgements
This material is based upon work supported by the 
National Science Foundation under Grants No. 
PHY1306520 (Nuclear Theory Program) and
PHY1452635 (Computational Physics Program).
We gratefully acknowledge discussions with L. Rammelm\"uller.

\appendix

\section{\label{Appendix1} Exact evaluation of the path integral for finite systems}

In order to illustrate the details as well as the generality of this technique, we evaluate the 
path integral for a four-component tensor from which each of the above traces may be obtained by suitable index contraction.  

To this end, we define
\beq
R_{ac,bd} = \int \mathcal D \sigma \;{\bd U}^{}_{}[\sigma]_{ab}\;{\bd U}^{}_{}[\sigma]_{cd}.
\eeq
We first write out each of the matrices ${\bd U}^{}_{}[\sigma]$ in its product form.  That is, we reintroduce the expression
\beq
{\bd U}^{}_{}[\sigma] = \prod_{j=1}^{N^{}_{\tau}}{\bd U}^{}_{j}[\sigma].
\eeq
For each contribution to the $N$-body transfer matrix, exactly $N$ factors of the matrix ${\bd U}^{}_{}[\sigma]$ appear, and as a result 
each temporal lattice point appears in the integrand $N$ times.  Writing out the integrand and grouping by timeslice, we obtain
\begin{widetext}
\bea
R^{}_{ac,bd} &=& \int \mathcal D \sigma \;{\bd U}^{}_{}[\sigma]_{ab}\;{\bd U}^{}_{}[\sigma]_{cd}
= \int \mathcal D \sigma \;\left({\bd U}^{}_{1}[\sigma]\;{\bd U}^{}_{2}[\sigma]\;\dots\;{\bd U}^{}_{N^{}_{\tau}}[\sigma]\right)_{ab}
\left({\bd U}^{}_{1}[\sigma]\;{\bd U}^{}_{2}[\sigma]\;\dots\;{\bd U}^{}_{N^{}_{\tau}}[\sigma]\right)_{cd} \\
&=& \sum\limits_{\substack{k_1,k_2,\dots,k_{N^{}_{\tau}-1} \\ l_1,l_2,\dots,l_{N^{}_{\tau}-1}}}\int \mathcal D \sigma \; 
\Big({\bd U}^{}_{1}[\sigma]_{a k_1}\;{\bd U}^{}_{1}[\sigma]_{c l_1}\Big)\;\Big({\bd U}^{}_{2}[\sigma]_{k_1 k_2}\;{\bd U}^{}_{2}[\sigma]_{l_1 l_2}\Big)\;\dots\;\Big({\bd U}^{}_{N^{}_{\tau}}[\sigma]_{k_{N^{}_{\tau}-1}b}\;{\bd U}^{}_{N^{}_{\tau}}[\sigma]_{l_{N^{}_{\tau}-1}d}\Big)\\
&=& \sum\limits_{\substack{k_1,k_2,\dots,k_{N^{}_{\tau}-1} \\ l_1,l_2,\dots,l_{N^{}_{\tau}-1}}}
\prod_{j=1}^{N^{}_{\tau}}\left(\int{\mathcal D\sigma(\tau_j)}\;{\bd U}^{}_{j}[\sigma]_{k_{j-1} k_{j}}\;{\bd U}^{}_{j}[\sigma]_{l_{j-1} l_{j}}\right),
\eea
\end{widetext}
where we set $k_0 = a$, $l_0 = c$, $k_{N_\tau} = b$, and $l_{N_\tau} = d$, and used the notation
\beq
\mathcal D\sigma(\tau) \equiv \prod_{{\bf r}} \frac{d \sigma({\bf r},\tau)}{2\pi}.
\eeq

Using the specific form of the individual $\bf U$ factors, we find
\bea
&&\!\!\!\!\!\!\!\int{\mathcal D\sigma(\tau_j)}{\bd U}^{}_{j}[\sigma]_{k_{j-1} k_{j}}{\bd U}^{}_{j}[\sigma]_{l_{j-1} l_{j}}\\
&=&\!\!\sum\limits_{\substack{p,q \\ p',q'}}\!\int{\mathcal D\sigma(\tau_j)}\!\!\left({\bd T}_{k_{j-1} p}{\bd V}^{}_{j}[\sigma]_{p q}{\bd T}_{qk_{j} }\right)\!\!\left({\bd T}_{l_{j-1} p'}{\bd V}^{}_{j}[\sigma]_{p' q'}{\bd T}_{q'l_{j} }\right), \nonumber
\eea
which using our chosen form of $\bf V$ becomes
\bea
&=&\!\!\sum\limits_{\substack{p,q \\ p',q'}} {\bd T}_{k_{j-1} p}{\bd T}_{qk_{j} }{\bd T}_{l_{j-1} p'}{\bd T}_{q'l_{j} }\delta^{}_{pq}\delta^{}_{p'q'}\times 
\nonumber\\
&&\int{\mathcal D\sigma(\tau_j)}\left(1 + A\;\sin\sigma(p,\tau^{}_{j})\right)\left(1 + A\;\sin\sigma(p',\tau^{}_{j})\right) 
\nonumber\\
&=&\!\!\sum\limits_{\substack{p,q \\ p',q'}} {\bd T}_{k_{j-1} p}{\bd T}_{qk_{j} }{\bd T}_{l_{j-1} p'}{\bd T}_{q'l_{j} }\delta^{}_{pq}\delta^{}_{p'q'} \left(1 + (e^{\tau g}-1)\delta^{}_{pp'}\right),\nonumber
\eea
where we used
\bea
\int{\mathcal D\sigma(\tau_j)}&& \!\!\!\!\!\!\!\!\!\!\!\! \left(1 + A\;\sin\sigma(p,\tau^{}_{j})\right)\left(1 + A\;\sin\sigma(p',\tau^{}_{j})\right)
\nonumber \\
&=& 
\left(1 + (e^{\tau g}-1)\delta^{}_{pp'}\right).
\eea

Thus, we arrive naturally at the definition 
\bea
[M^{}_{2}]^{}_{ac,bd} &=& {\mathcal K}_{ab} {\mathcal K}_{cd} + (e^{\tau g}-1) {\mathcal I}_{abcd},
\eea
as the transfer matrix in the two-particle subspace, where
\bea
{\mathcal K}_{ij} &=& \sum_{p}{\bd T}_{i p}{\bd T}_{pj}, \\
{\mathcal I}_{ijkl} &=& \sum_{p}{\bd T}_{i p}{\bd T}_{pj }{\bd T}_{k p}{\bd T}_{pl}.
\eea

Indeed, this definition of $M_2$ as a transfer matrix makes sense, because
\beq
R^{}_{ac,bd} =  \sum\limits_{\substack{k_1,k_2,\dots,k_{N^{}_{\tau}-1} \\ l_1,l_2,\dots,l_{N^{}_{\tau}-1}}} 
\; \prod_{j=1}^{N^{}_{\tau}}[M^{}_{2}]_{k_{j-1} k_j,l_{j-1} l_j},
\eeq
or more succinctly, 
\beq
R^{}_{ac,bd} = \left [M^{N^{}_{\tau}}_{2} \right]_{ac,bd}.
\eeq

In a similar fashion, one may show without much difficulty that the transfer matrix of the three-body problem
(for distinguishable particles, i.e. no symmetrization or antisymmetrization is enforced) is
\bea
[M^{}_{3}]^{}_{abc,def} = {\mathcal K}_{ad} {\mathcal K}_{be} {\mathcal K}_{cf} + (e^{\tau g}-1) {\mathcal J}_{abc,def},
\eea
where
\bea
{\mathcal J}_{ijk,lmn} = {\mathcal K}_{il} {\mathcal I}_{jkmn} + {\mathcal K}_{jm} {\mathcal I}_{ikln} + {\mathcal K}_{kn} {\mathcal I}_{ijlm}.
\eea
The pattern from this point on is clearly visible: there is one term for each `spectator' particle that does not participate in the interaction, while
the other two are accounted for by an interacting term governed by the $\mathcal I_{abcd}$ object. One may thus infer the form of the transfer
matrix for higher particle numbers.
\\

\section{\label{Appendix2} Auxiliary parameter dependence}

In this Appendix we show a few more examples on the mild dependence of the
entanglement-entropy derivative $\langle\ln{Q}[{\bd \sigma}]\rangle_\lambda^{}$
as other parameters are varied. In all cases, the data shown corresponds to
full 3D calculations in the unitary regime.

\begin{figure}[t]
\includegraphics[width=1.0\columnwidth]{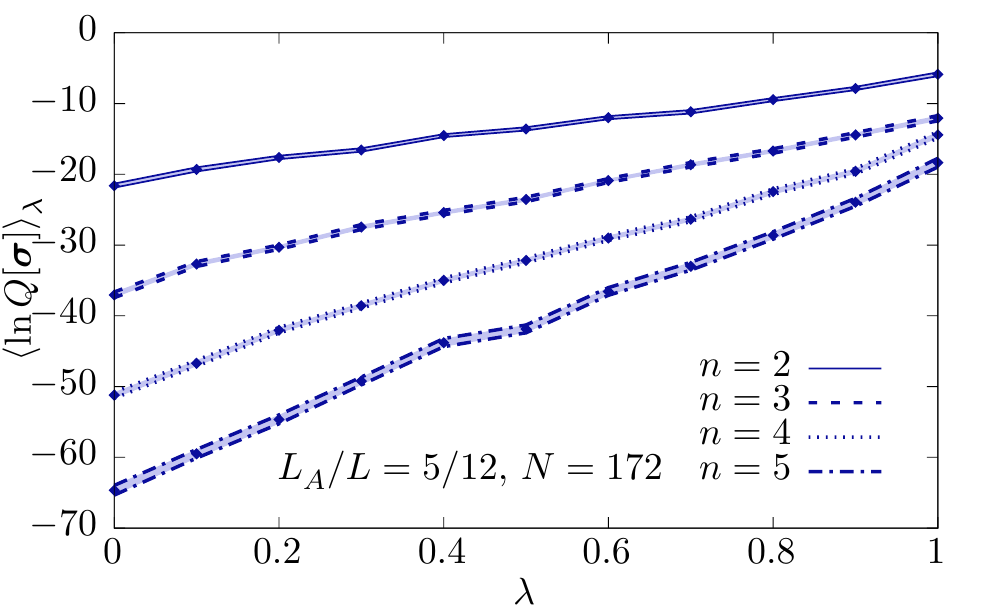}
\includegraphics[width=1.0\columnwidth]{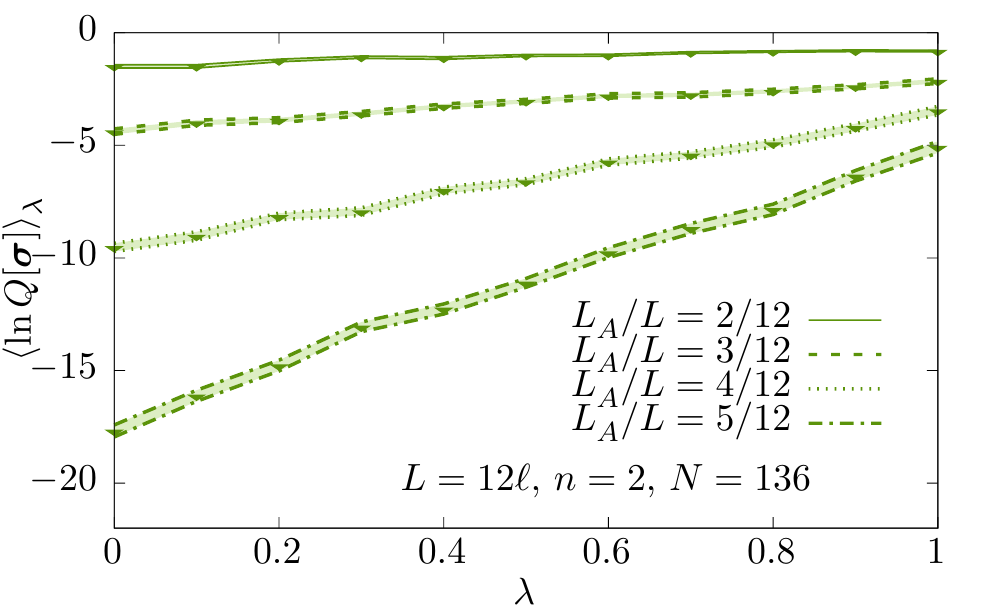}
\caption{\label{Fig:EEIOrderSections} 
Top: $\lambda$ dependence of $\langle\ln{Q}[{\bd \sigma}]\rangle_\lambda^{}$ for several 
R\'enyi orders $n=2,3,4,5$, for subsystem size $L_A= 5/12 L$,
for $N=172$ fermions at unitarity in a box of size $L=N_x\ell$ (where $N_x=12$ points and $\ell=1$).
Bottom: $\lambda$ dependence of $\langle\ln{Q}[{\bd \sigma}]\rangle_\lambda^{}$ for several subsystem sizes $L_A$,
for $N=136$ fermions at unitarity in 
a box of size $L=N_x\ell$ (where $N_x=12$ points and $\ell=1$), and for R\'enyi order $n=2$.
}
\end{figure}

In Fig.~\ref{Fig:EEIOrderSections} (top) we show the variation of that derivative when
the R\'enyi order is changed from $n=2$ to $n=5$, at fixed particle number and region size.
In Fig.~\ref{Fig:EEIOrderSections} (bottom) we show how $\langle\ln{Q}[{\bd \sigma}]\rangle_\lambda^{}$
changes when the particle number is varied, at fixed R\'enyi order $n$.
%


\end{document}